\documentclass[]{aa}

\usepackage{amssymb}
\usepackage{hyperref}
\usepackage[usenames, dvipsnames]{color}
\usepackage{graphicx}
\usepackage{natbib}
\usepackage[rightcaption]{sidecap}
\usepackage{txfonts}
\usepackage{units}
\usepackage{url}
\usepackage{pifont}
\usepackage[modulo, switch]{lineno}

\definecolor{blue}{rgb}{0.00, 0.00, 1.00}
\definecolor{red}{rgb}{0.86, 0.08, 0.24}
\definecolor{orange}{rgb}{1.00, 0.55, 0.00}
\definecolor{darkblue}{rgb}{0.00, 0.00, 0.55}
\definecolor{green}{rgb}{0.00, 0.39, 0.00}
\definecolor{pink}{rgb}{1.000000,0.078431,0.576471}

\newcommand\arcdeg{\mbox{$^\circ$}}

\definecolor{myDarkRed}{rgb}{0.698, 0.094, 0.133}

\begin{document} 


\title{Temporal evolution of arch filaments as seen in \ion{He}{i} 10830~\AA}
\titlerunning{Temporal evolution of arch filaments as seen in \ion{He}{i} 10830~\AA}

\author{%
    S.J.\ Gonz{\'a}lez Manrique\inst{1,2,3}, 
    C.\ Kuckein\inst{2}, 
    M.\ Collados\inst{4}, 
    C.\ Denker\inst{2},
    S.K.\ Solanki\inst{5, 6},
    P.\ G{\"o}m{\"o}ry\inst{1}, \\
    M.\ Verma\inst{2},
    H.\ Balthasar\inst{2},
    A.\ Lagg\inst{5}, and
    A.\ Diercke\inst{2,3} 
    }
\authorrunning{Gonz{\'a}lez Manrique et al.}
   
\institute{%
    $^1$ Astronomical Institute, Slovak Academy of Sciences, 
         05960 Tatransk\'{a} Lomnica, Slovak Republic\\
    $^2$ Leibniz-Institut f{\"u}r Astrophysik Potsdam (AIP),
         An der Sternwarte 16, 
         14482 Potsdam, Germany\\
    $^3$ Institut f{\"u}r Physik and Astronomie, Universit{\"a}t Potsdam, 
         Karl-Liebknecht-Stra{\ss}e 24/25,
             14476 Potsdam-Golm, Germany\\
    $^4$ Instituto de Astrof{\'i}sica de Canarias,
             c/ V{\'i}a L{\'a}ctea s/n, 
             38205 La Laguna, Tenerife, Spain\\
    $^5$ Max-Planck-Institut f{\"u}r Sonnensystemforschung,
         Justus-von-Liebig-Weg 3,
         37077 G{\"o}ttingen, Germany\\
    $^6$ School of Space Research, Kyung Hee University,
         Yongin, Gyeonggi-Do, 446-701, Republic of Korea\\                            
    \email{smanrique@ta3.sk}}

\date{Received January 22, 2018; accepted June 18, 2018}
 
\abstract
{}
{We study the evolution of an arch filament system (AFS) and of its individual 
arch filaments to learn about the processes occurring in them.}
{We observed the AFS at the GREGOR solar telescope on Tenerife at high cadence 
with the very fast spectroscopic mode of the GREGOR Infrared Spectrograph 
(GRIS) in the \ion{He}{i} 10830~\AA\ spectral range. 
The \ion{He}{i} triplet profiles were fitted with analytic functions 
to infer line-of-sight (LOS) velocities to follow plasma motions within the AFS.}
{We tracked the temporal evolution of an individual arch filament
over its entire lifetime, as seen in the \ion{He}{i} 10830~\AA\ triplet. 
The  arch filament expanded in height and extended in
length from 13\arcsec\ to 21\arcsec. The lifetime of this arch filament is about 30 min.
About 11~min after the arch filament is seen in \ion{He}{i}, the loop
top starts to rise with an average Doppler velocity of 6~km~s$^{-1}$. Only two minutes later, 
plasma drains down with supersonic velocities towards the footpoints reaching a peak velocity of up to
40~km~s$^{-1}$ in the chromosphere. The temporal evolution of 
\ion{He}{i} 10830~\AA\ profiles near the leading pore showed almost 
ubiquitous dual red components of the \ion{He}{i} triplet,
indicating strong downflows, along with material nearly at rest 
within the same resolution element during the whole observing time.}
{We followed the arch filament as it carried plasma during its 
rise from the photosphere to the corona. The material then drained toward 
the photosphere, reaching supersonic velocities, 
along the legs of the arch filament. Our observational results support 
theoretical AFS models and aids in improving future models. 
}

    
\authorrunning{Gonz{\'a}lez Manrique et al.}

\maketitle
   

\section{Introduction}\label{SEC1}

An important role in solar physics in general, and in solar activity in
particular, is played by the emergence of magnetic flux at the solar surface. 
The generally accepted theory
states that emerging flux regions (EFRs) are formed by magnetic flux tubes 
that are transported to the solar surface by buoyancy \citep{Parker1955, Zwaan1987, Caligari1995}.
The process forms $\Omega$-loops and can lead to active regions. After the
flux finishes rising to the photosphere, the active regions start to decay
\citep{vanDrielGesztelyi2015}. 

An EFR as seen in the photosphere can evolve on many different temporal and spatial scales.
The smallest EFRs have a lifetime of just a few 
minutes, while large active regions have a lifetime from days up to months
\citep{Archontis2012}. 

Many different magnetic field elements emerge or disappear
on the solar surface when EFRs form. In most cases, these elements
evolve into bipolar structures. Smaller magnetic elements can appear and merge with
the main bipole or are canceled with opposite polarity elements
\citep{Archontis2012}. In many EFRs the magnetic field between two opposite main
polarities can form the well known ``sea-serpent'' topology  
\citep[see, e.g., ][]{Harvey1973, Pariat2004, Archontis2012, Cheung2014,
Schmieder2015}. Emerging field lines can form a
photospheric pattern of aligned dark intergranular lanes \citep{Strous1996}.
In the chromosphere, EFRs become visible through magnetic
loops loaded with plasma, which flows down towards both footpoints \citep{Solanki2003b}.
In this study, we observed a chromospheric arch filament system (AFS) connecting two photospheric
pores (see Sect.~\ref{SEC2_4}). 

The magnetic activity in the solar photosphere is characterized by various magnetic
structures with a wide range of spatial sizes.
The magnetic radius of pores is always larger than the continuum radius, which implies
that a magnetic canopy exists \citep{Keil1999,Buehler2015}.

Velocities perpendicular to the surface and horizontal motions inside the concentrated magnetic fields of
mature pores are generally close to zero for all photospheric heights, and the 
observed strong downflows in the surroundings  range between 0.5 and 3~km~s$^{-1}$
\citep[see, e.g., ][]{Brants1985,Leka1998,Hirzberger2003, Sankarasubramanian2003, Cho2010, Sobotka2012, QuinteroNoda2016}.
\citet{QuinteroNoda2016} measured small upward velocities up to 1~km~s$^{-1}$ at mid-photospheric 
layers. \citet{Verma2014} calculated the horizontal velocities of more than 2800 pores.
Isolated pores exhibit minor inflows in the interior and stronger outflows surrounding
the pores, which reach up to 0.7~km~s$^{-1}$.

The chromosphere above an EFR is best described as an
AFS, which connects two regions of opposite magnetic polarity crossing 
the polarity inversion line.
\citet{Bruzek1967} reported these systems of fibrils for the 
first time and called them arch filament systems. The AFSs are prominently visible in line-core 
filtergrams of the strong chromospheric absorption line H$\alpha$ and also
in the line core of the \ion{Ca}{ii} \,H \&\,K lines \citep{Bruzek1969}.
More recently they were observed in the \ion{He}{i} 10830~\AA\ triplet
\citep[e.g.,][]{Solanki2003b, Spadaro2004, Lagg2004, Lagg2007, Xu2010, VargasDominguez2012,Ma2015, GonzalezManrique2016}. \citet{Bruzek1969} determined that the length of the arch 
filaments reaches the size of supergranular network cells (20\,--\,30~Mm). 
Typically, the width of individual AFS loops is only a few megameters, and the 
height of the arches varies between 5 and 15~Mm \citep{Merenda2011} with a 
lifetime of about 30~min \citep{Bruzek1967}. However, some individual loops of the
AFS can reach heights close to 25~Mm and lengths of 20\,--\,40~Mm
\citep{Tsiropoula1992}. \citet{Solanki2003b} reported upflows in the center of 
the arches and downflows at the footpoints. Downflows reach velocities in the 
range of 30\,--\,50~km~s$^{-1}$. They were observed near both footpoints of 
the AFS, whereas loop tops rise with about 1.5\,--\,20~km~s$^{-1}$ 
\citep[see, e.g.,][]{Bruzek1969, Zwaan1985, Chou1988, Lites1998, Solanki2003b, Spadaro2004, GonzalezManrique2017b}. 

A suitable spectral line to investigate chromospheric features and particularly AFSs is the 
\ion{He}{i} at 10830\AA\ triplet \citep{Lagg2007, Xu2010}. It
forms in the upper chromosphere \citep{Avrett1994} and consists of transitions between the
upper levels $2^{3}P_{2, 1, 0}$ and the lower level $2^{3}S_{1}$. Two of the
transitions overlap and consequently only two spectral lines can be observed. The
nonblended line observed at 10829.09~\AA\ is commonly called the ``blue'' component
and the blended line at 10830.30~\AA\ is called the ``red'' component. 
The exact wavelength positions were taken from the National Institute of Standards 
and Technology (NIST).\footnote{www.nist.gov}

Several studies of the chromospheric dynamics have been carried out using the 
\ion{He}{i} triplet. Particularly,
very high downflow velocities were reported \citep[e.g.,][]{penn1995,muglach1997,muglach1998}. 
The observations very often show that two 
or more atmospheric components are present within the
same spatial resolution element. 
It is possible that such complex, multicomponent  profiles
provide information of two or more different heights in the atmosphere.
\citet{Schmidt2000b} reported for the first time these ``dual or multiple flows,''
 indicated by two or more peaks in the spectrum close to the \ion{He}{i} triplet. 
In the case of only two components, it is common to find one of the peaks with 
subsonic velocities, called the ``slow component,'' while the other peak reaches 
supersonic velocities and is called the ``fast component.''  Flow speeds above 10\,km\,s$^{-1}$ in
the \ion{He}{i} triplet are considered to be supersonic
\citep{AznarCuadrado2005,AznarCuadrado2007}. Studies of AFS based on the \ion{He}{i}
triplet reported downflows of 15\,--\,90~km~s$^{-1}$ 
\citep[see, e.g.,][]{Solanki2003b, AznarCuadrado2005,  AznarCuadrado2007, Sasso2007, Lagg2007, Balthasar2016, GonzalezManrique2016}. 
Results of the temporal evolution
of dual-component \ion{He}{i} 10830~\AA\ spectral profiles were 
presented in \citet{GonzalezManrique2017}.

\citet{Solanki2003b} investigated the magnetic field in an AFS, 
and reported magnetic field strengths in
the footpoints between 390 and 500~G at chromospheric heights (as deduced from the \ion{He}{i} triplet),
the strength decreasing to 50~G at the tops of the arch filaments.
Magnetic field strengths between 700 and 900~G were reported by
\citet{Lagg2007} in regions with very high velocities, typically located at the
footpoints. \citet{Xu2010} inferred the magnetic field strength
showing an asymmetric distribution from the footpoint (800~G) to the loop tops (300~G).
 \citet{Solanki2003b}, 
and \citet{Xu2010} reconstructed the three-dimensional magnetic field in
an AFS. They assumed that the \ion{He}{i} 10830~\AA\ triplet is formed along the
magnetic field loops, and \citet{Merenda2011} confirmed this assumption.

The new generation of solar telescopes and
instruments allow us to record  very high-resolution observations necessary to
investigate the dynamics, magnetic field, and widths of arch filaments.
These observations will help us to
answer many open questions related to flux emergence
\citep[see the questions raised in the review by ][]{Cheung2014}: (1) What are the
observational consequences of the emerging flux \citep{Bruzek1967,Zwaan1978}? (2) How
do EFRs evolve with time in the different layers of the solar atmosphere and how are these layers
linked? (3) Is it possible to measure the height difference between the photosphere
and the chromosphere connected by the legs of the AFSs? 

In Sect.~\ref{SEC2_4} we present the observations. We describe image restoration
and image processing  in Sect.~\ref{SEC3_4}. In Sect.~\ref{SEC4_4} we introduce 
the data analysis. In Sect.~\ref{SEC5_4} we present the results, and analyze mainly the
temporal evolution of the AFS using the unique observations of the very fast spectroscopic mode 
of GRIS (see Sect.~\ref{SEC2_4}) in the \ion{He}{i} 10830~\AA\ spectral region. Finally, we discuss 
our findings and compare them with the more recent literature concerning the 
AFSs and EFRs in Sect.~\ref{SEC6}. Some initial results of this study were
presented in \citet{GonzalezManrique2016, GonzalezManrique2017}.

%
%

\section{Observations}\label{SEC2_4}

\begin{figure}[t]
\includegraphics[width=\columnwidth]{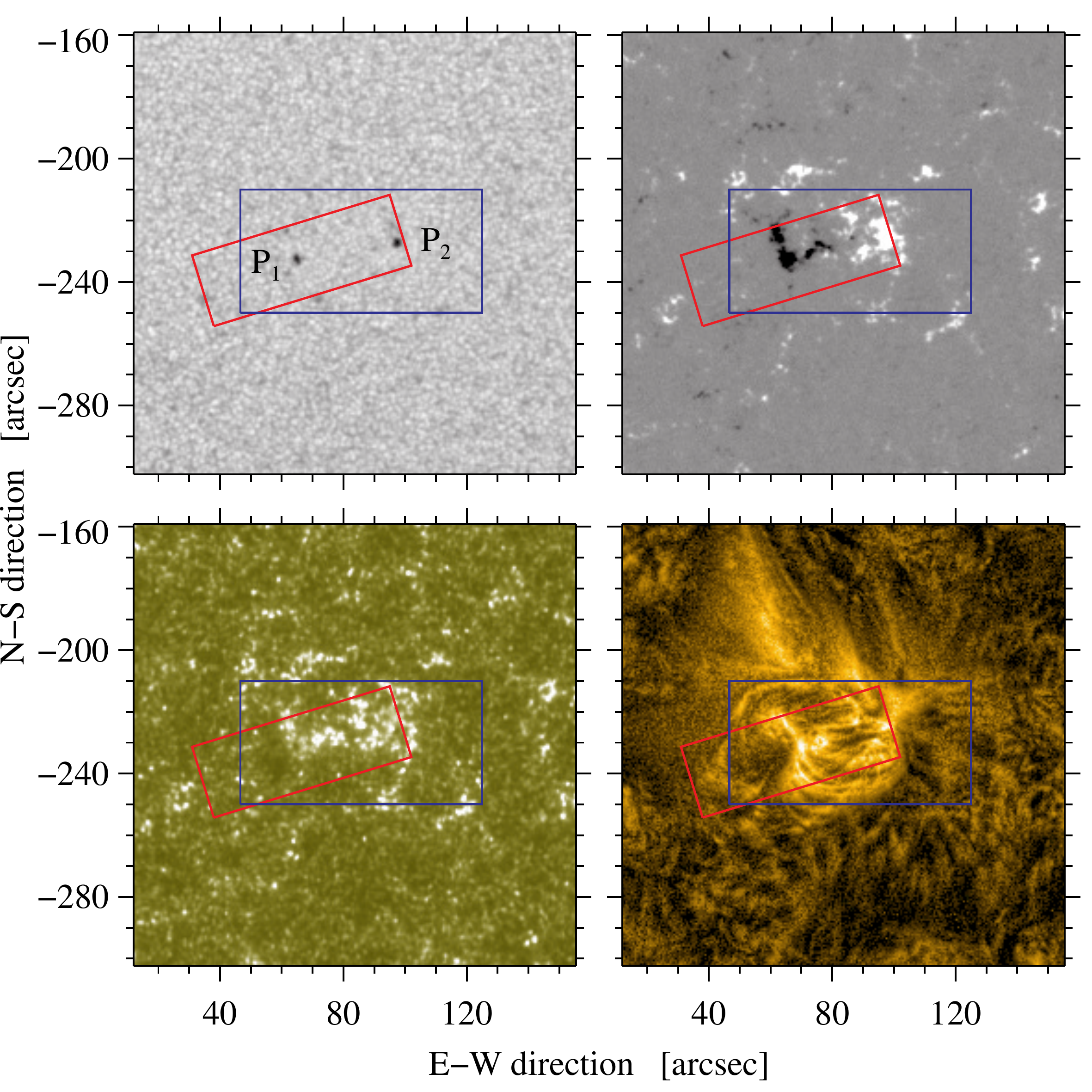}
\caption{Overview of the EFR at 09:00~UT on 2015 April~17: SDO HMI continuum 
    image (\textit{top left}), HMI magnetogram (\textit{top right}), AIA
    1600~\AA\ image (\textit{bottom left}), and NAFE-corrected AIA
    \ion{Fe}{ix} 171~\AA\ image (\textit{bottom right}). The red
    rectangle delineates the FOV covered by GRIS. The blue rectangle
    represents the FOV where the temporal evolution of the magnetic
    flux was calculated (see Fig.~\ref{FIG05_4}).}
\label{FIG01_4}
\end{figure}

An EFR smaller than a medium-sized active region, containing two principal 
pores with opposite polarities and an associated AFS in the chromosphere, 
was observed between 08:16~UT and 09:20~UT on 2015 April~17. The
region of interest (ROI) is located at heliographic coordinates S19 and W4
($\mu \equiv \cos\theta = 0.97$). The ground-based observations were 
performed with the GREGOR Infrared Spectrograph \citep[GRIS,][]{Collados2012}
located at the 1.5-meter GREGOR solar telescope \citep{Schmidt2012, 
Denker2012} at Observatorio del  Teide, Tenerife, Spain. The overview of 
the EFR/AFS in Fig.~\ref{FIG01_4} contains the field of view (FOV) of the
high-resolution observations with GRIS, which is superposed on extreme
ultraviolet (EUV) and continuum images and a photospheric magnetogram 
obtained with the Atmospheric Imaging Assembly \citep[AIA,][]{Lemen2012} 
and the Helioseismic and Magnetic Imager \citep[HMI,][]{Scherrer2012, 
Schou2012}, respectively, both on board the Solar Dynamics Observatory
\citep[SDO,][]{Pesnell2012} at 09:00~UT on 2015 April~17. 

GRIS was operated in the very fast spectroscopic mode (no polarimetry) in the
\ion{He}{i} 10830~\AA\ spectral region \citep{GonzalezManrique2017}. With this
fast mode, it is possible to scan the ROI at a much higher cadence with only
a single exposure per slit position while accepting a higher noise level 
than in the standard polarimetric mode, which typically requires several
accumulations for each polarization state. In addition, polarimetric
information was purposefully discarded, again to achieve the high cadence
required to investigate very dynamic processes in the chromosphere.
Preliminary results obtained from this data set discussing
noise level and fitting procedures for dual-component \ion{He}{i} 10830~\AA\
spectral profiles were presented in \citet{GonzalezManrique2016}.

\begin{figure}[t]
\includegraphics[width=\columnwidth]{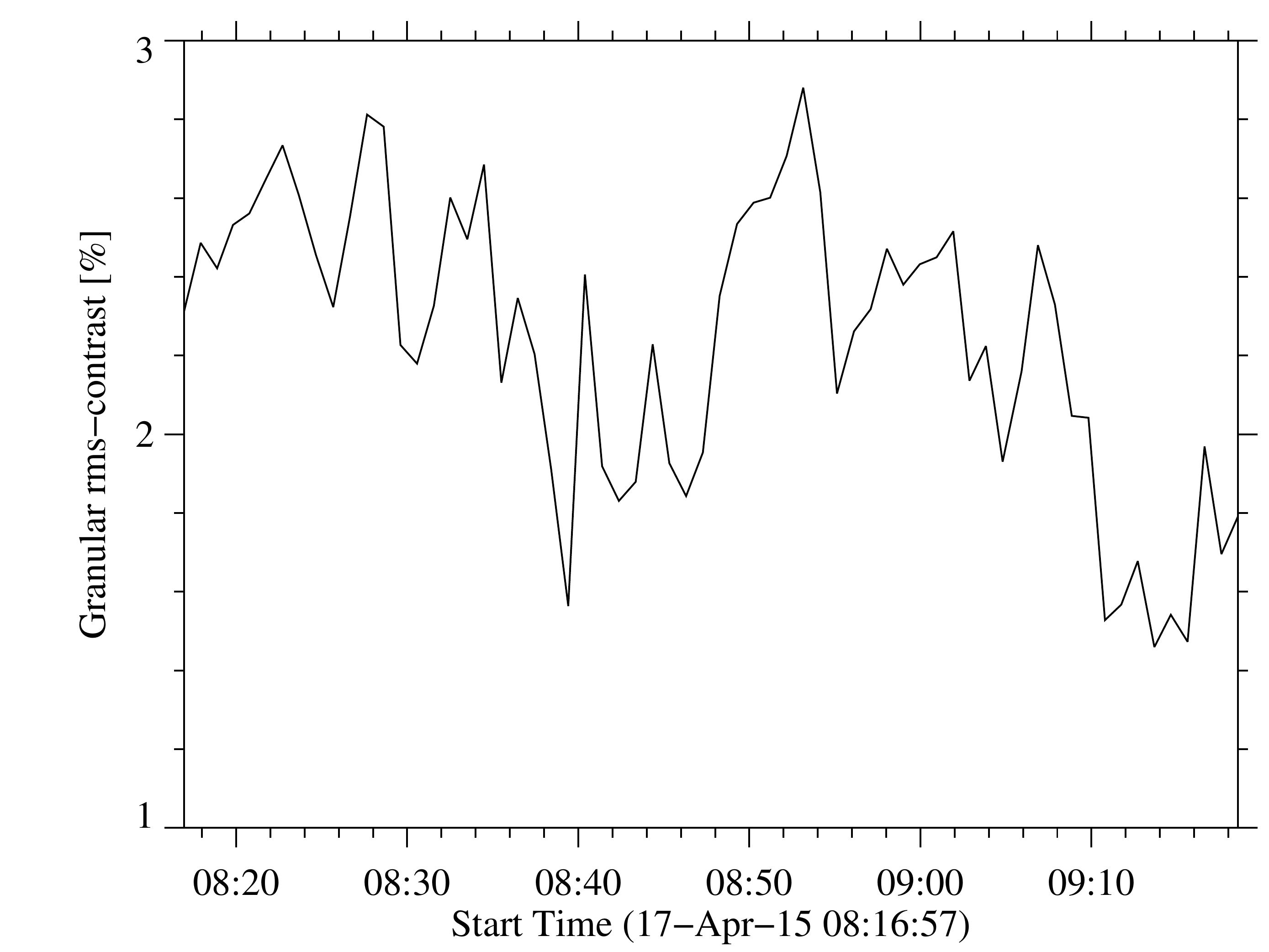}
\caption{Temporal evolution of the granular rms-contrast during the observing
    time with GRIS. In all figures the observing period refers to universal
    time.}
\label{FIG_CONTRAST}
\end{figure}

\begin{figure}[t]
\includegraphics[width=\columnwidth]{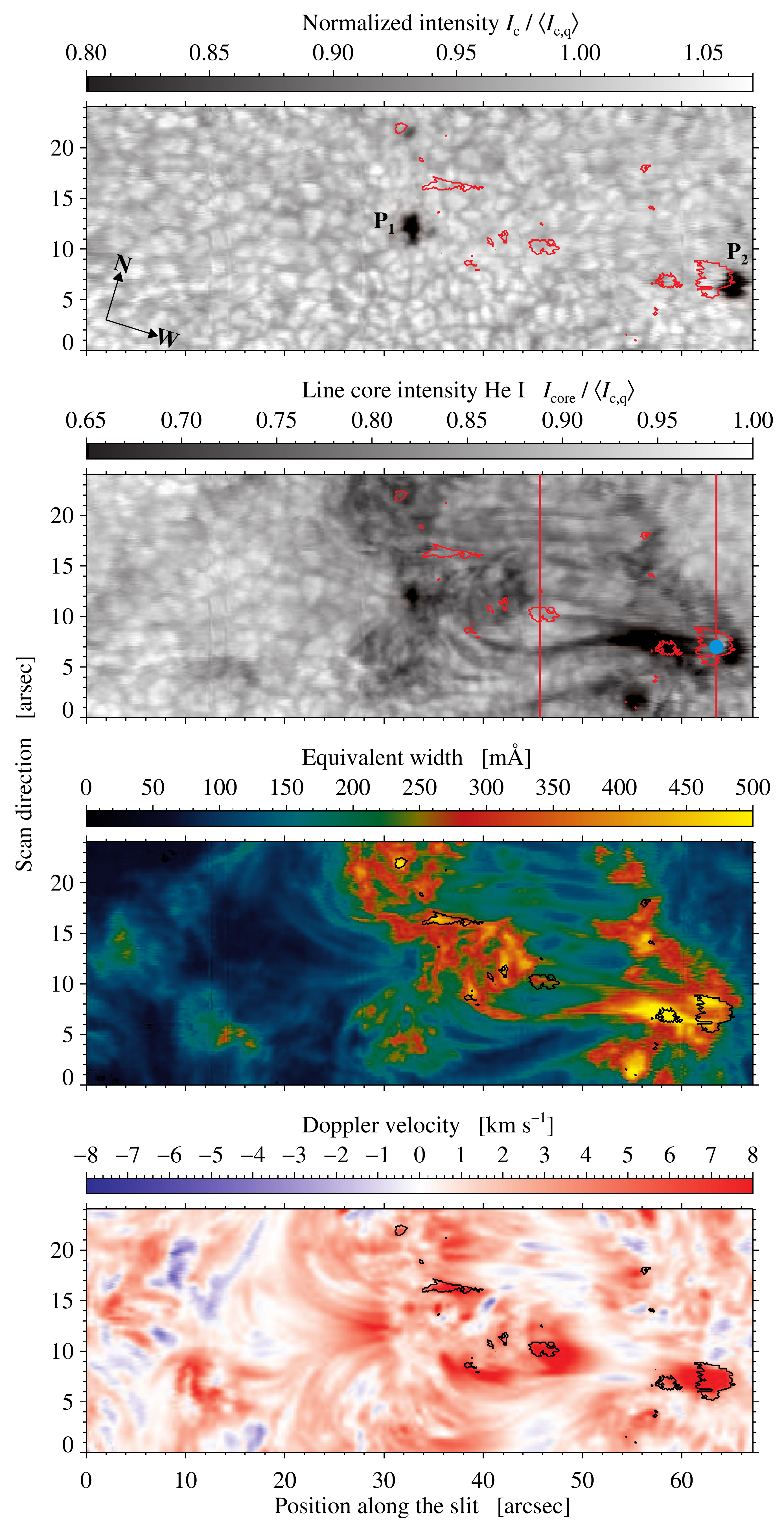}
\caption{Slit-reconstructed GRIS images at 09:00:56~UT on 2015 April~17 of the 
   EFR centered on the pore P$_1$ with negative polarity shown in Fig.~\ref{FIG01_4}:
   continuum intensity, line core intensity of the red component of the 
   \ion{He}{i} triplet, equivalent width, and \ion{He}{i} Doppler velocity
   calculated assuming only a single flow component (\textit{top to bottom}). 
   The blue filled circle (\textit{second panel from top}) refers to the location of strong
   downflows near the pore P$_2$, and it is used to calculate the average
   profiles in Fig.~\ref{HE_PROFILES}. The contours encompass dual-flow 
   \ion{He}{i} profiles. The red vertical lines indicate positions
   of the space-time diagrams (the right line 
   corresponds to the upper panel of Fig.~\ref{FIG_EVOL_INT}). The temporal
   evolution of the He\,\textsc{i} line depression and Doppler shifts is
   available as an online movie covering the period between 08:16:57~UT and
   09:18:33~UT.} 
\label{FIG02_4}
\end{figure}

The observed spectral region covers a wavelength range in the near-infrared
(NIR) of about 18~\AA\ containing the photospheric \ion{Si}{i} 10827~\AA,
\ion{Ca}{i} 10834~\AA, and \ion{Ca}{i} 10839~\AA\ lines; the chromospheric
\ion{He}{i} 10830~\AA\ triplet; and other solar and telluric lines. The
dispersion is $\delta_{\lambda} = 18.0$~m\AA\
pixel$^{-1}$, and the number of spectral  points along the wavelength axis is 
$N_{\lambda} = 1010$. The spatial step  size and the pixel size along the 
slit are very similar,  0$\farcs$134  and 0$\farcs$136, respectively. 
The integration time is $t = 100$~ms for one accumulation. A spatial scan 
consists of 180 steps in total. Thus, the FOV is $66\farcs3 \times 24\farcs1$,
and it takes about 58~s to cover it. This time interval includes overhead 
for reading out the camera and writing the data to disk. The observations
continued for about one hour resulting in 64 spatio-spectral data cubes. The
NIR spectra were taken under good seeing conditions. As shown in
Fig.~\ref{FIG_CONTRAST}, the seeing changed during the time series, with
extended periods of very good observing conditions. This is quantified by
computing the rms-contrast of the granulation for each map. The values
fluctuated between 1.5\% and 2.9\% for the continuum at 1.0~$\mu$m 
during the observing period. As a comparison, previous works found 
an rms-contrast at disk center of 6.1\% and 2.9\% for 0.8~$\mu$m and
1.5~$\mu$m, respectively \citep{SanchezCuberes2003}. However, these values
refer to imaging data with 2\,--\,4 times shorter exposure times. At around
09:00~UT the seeing conditions were good to very good with a granular
rms-contrast between 2.2\%\ and 2.9\%.
Therefore, the $45\mathrm{th}$ scan at 
09:00:56~UT (see Fig.~\ref{FIG02_4}) was selected as a reference for the GRIS
data. 
The GREGOR Adaptive Optics System \citep[GAOS,][]{Berkefeld2012} provided 
real-time correction, ensuring a better quality of the spectral data cubes.
GREGOR's alt-azimuthal mount \citep{Volkmer2012} introduced an image rotation
of $22.3\arcdeg$, which had to be corrected in the data reduction shrinking
significantly the common FOV for the whole time series.

Context data are provided by SDO, where HMI takes full-disk filtergrams.
The filtergrams are combined to form continuum intensity images, 
LOS magnetograms, and Dopplergrams at a 45~s cadence, and 
the image scale is 0\farcs5 pixel$^{-1}$. The longitudinal 
magnetograms are obtained with a precision of
10~G and allow us to track the temporal evolution of the magnetic flux 
contained within the active region. The response of upper atmospheric layers,
i.e., the transition region and corona, is monitored at different plasma
temperatures with EUV and UV images provided by AIA at a cadence of 12~s and
24~s, respectively. The image scale of 0\farcs6 pixel$^{-1}$ is slightly 
larger  compared to HMI. The EUV
images are beneficial for establishing the topology of low-lying filamentary
structures and overarching loops, which are indicative of the magnetic
connectivity within the active region and its surroundings. 

%
%

\section{Data reduction}\label{SEC3_4}

Reduction and calibration of the GRIS data cube includes standard dark and
flat-field corrections \citep{GonzalezManrique2016}. Wavelength
calibration comprises orbital motions and solar gravity redshift corrections, 
as described by \citet[][]{Kuckein2012b}. Hence, the Doppler velocities refer
to an absolute scale. Common fringes, dust particles along the slit, and
abnormal intensity peaks are removed, and the proper continuum of the spectral
line is determined. In addition, the telluric line at 10832.108~\AA\ was
eliminated from the spectra \citep{GonzalezManrique2016}. This telluric line
interferes in some cases with the fast component of the \ion{He}{i} 10830~\AA\
triplet. To remove the telluric line, every profile was normalized to
the local continuum to ensure that the continuum of the telluric line
conforms to I$_c$ = 1. This normalization was only used to calculate the
Doppler shifts explained in Sect.~\ref{SEC4_4.3} and remove the telluric line.
We computed a mean intensity profile for 
every map, assuming that the central wavelength of the telluric line
is constant. We created a synthetic telluric profile by fitting the observed 
mean profile with a single Lorentzian profile. Subsequently, we divided every
spectral profile of the map by the synthetic profile, effectively removing the
telluric line. The slit-reconstructed continuum and the line-core intensity
map of the red \ion{He}{i} component are depicted in the first and second
panel of Fig.~\ref{FIG02_4}, respectively.

\begin{figure}[t]
\includegraphics[width=\columnwidth]{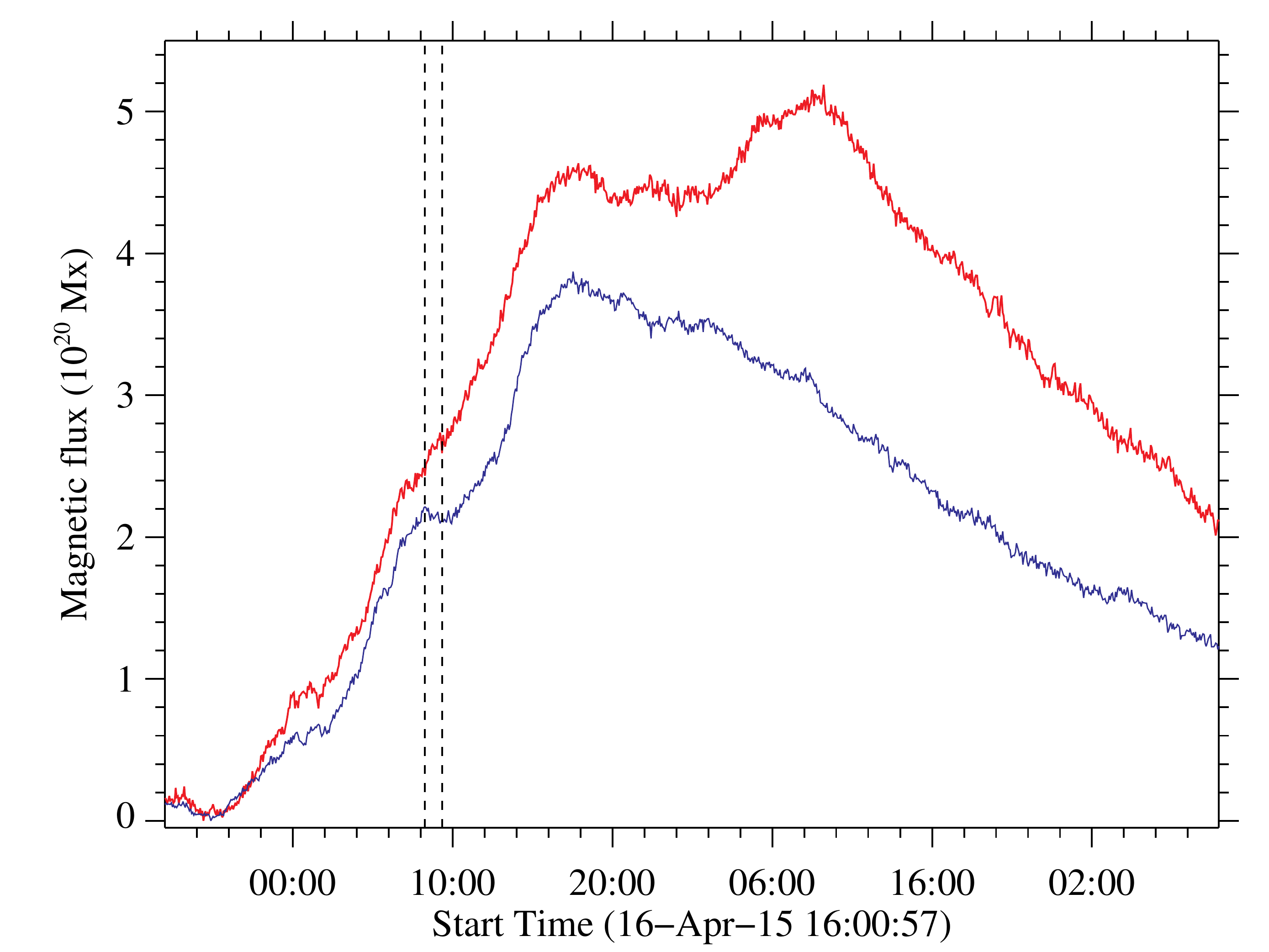}
\caption{Temporal evolution of new magnetic flux from HMI magnetograms 
   covering the entire ROI (blue rectangle in Fig.~\ref{FIG01_4}). The
   red and blue solid lines represent the magnetic flux of the positive and
   negative polarity, respectively. The dashed vertical lines mark start and 
   end of the high-resolution observations with GREGOR.}
\label{FIG05_4}
\end{figure}

The AIA and HMI data were downloaded from the Virtual Solar Observatory (VSO) 
in level 1.0 and 1.5 format, respectively. The different AIA filtergrams and 
HMI data products are adjusted to a common coordinate system and image scale
(0\farcs6 pixel$^{-1}$).
All data were compensated for differential rotation with respect to the central 
meridian. The reference image was taken at 00:00:57~UT on 2015 April~17 
because at this time the EFR was exactly at the central meridian. The
\ion{Fe}{ix} 171~\AA\ image in the bottom right panel of Fig.~\ref{FIG01_4} 
is enhanced using the Noise Adaptive Fuzzy Equalization method 
\citep[NAFE, ][]{Druckmueller2013}.

%
%

\section{Data analysis}\label{SEC4_4}


\subsection{Temporal evolution of emerging magnetic flux}
\label{SEC4_4.1}

Our study of the temporal evolution of the emerging magnetic flux is based on time series  
of HMI magnetograms compensated for solar differential rotation. 
The considered HMI data cover both flux emergence and decay. 
The high-resolution GRIS data were recorded 
roughly in the middle of the flux emergence period (see vertical dashed lines 
in Fig.~\ref{FIG05_4}). To compute the magnetic 
flux, a blue rectangular ROI was selected encompassing the entire EFR (Fig.~\ref{FIG02_4}). 
The ROI extends  in disk center coordinates from the  bottom left corner
($+45\arcsec$, $-250\arcsec$) to the top right corner ($+125\arcsec$, $-210\arcsec$).
The rise and decay of the magnetic flux is tracked for both magnetic
polarities are shown in Fig.~\ref{FIG05_4}, where the red and blue curves
correspond to the positive and negative magnetic polarities, respectively. The
pre-existing magnetic flux was determined as the median value of a two-hour
time series just before onset of the flux emergence at 16:00~UT on 2015 April~16
and was subtracted separately for each polarity; i.e.,  we only estimated the 
total amount of newly emerging flux.


\subsection{Line-of-sight velocities of the \ion{He}{i} 10830~\AA\ triplet}
\label{SEC4_4.3}

Generally, the \ion{He}{i} 10830~\AA\ triplet in the observed data cube 
only consists  of  one blue and two blended red components. Nevertheless, a
small percentage of spectral profiles reveals clear signatures of ``dual
flows'' \citep{Schmidt2000b}, which splits each blue and red component into 
a slow and fast  component (for a total of 4). The centroid of the slow component is
commonly close to rest, while the centroid of the fast component is in most
cases strongly redshifted. An approach to fit the two parts of the red
component is described by \citet{GonzalezManrique2016} based on a subset 
of the current data set. Initially, all profiles are fitted with a single
Lorentzian profile using MPFIT \citep{Markwardt2009}, a program written 
in the interactive data language (IDL). The wavelength range automatically 
adjusts to the line depths, i.e., it is broader for strong lines, to enhance
the accuracy and stability of the fits. The slow component of the \ion{He}{i}
red component is potentially blended with the fast component of the
\ion{He}{i} blue component. The random and systematic errors induced by this
method are small enough to allow scientific interpretation of the 2D maps of
physical parameters. The wavelength reference for the LOS velocities is the
average laboratory wavelength 10830.30~\AA\ of the blended red component 
given by the NIST database. 

In a second step, only dual-flow profiles are fitted with a double-Lorentzian
profile. This deviates from the approach in \citet{GonzalezManrique2016}, where
all profiles were fitted with a double-Lorentzian followed by an evaluation,
whether single- or double-peaked profiles are more appropriate. In the current
implementation dual-flow profiles are automatically detected based on the
single-Lorentzian fits. Their locations are indicated by the red or black
contours in all panels of Fig.~\ref{FIG02_4}. The selection criteria for dual-flow
profiles are (1) a strong asymmetry between the blue and red line wings at half 
the line depth, (2) the presence of two clearly defined minima, (3) a large FWHM of the line
profiles, and (4) a strongly redshifted line core with respect to the mean
central wavelength.

%
%

\section{Results}
\label{SEC5_4}


\subsection{General description of the region of interest}\label{SEC5_4_1}

The high-resolution observations cover a short period in the EFR
lifetime, which is marked by dashed vertical lines in Fig.~\ref{FIG05_4}.
At the time when the observations started, roughly half the final flux
had already emerged (see Fig.~\ref{FIG05_4}). Thus, the outcome of this
study primarily addresses questions related to ongoing flux emergence.
The lifetime of the EFR,  as observed with the SDO data
(see Fig.~\ref{FIG01_4}), is about three days. The ROI has two main pores with opposite
polarities. 
This bipolar region shows many properties of an active region, but 
likely does not carry enough flux to develop mature sunspots. The 
two pores become visible in HMI magnetograms at around 20:08~UT on 
2015 April~16. In HMI continuum images, the pores appear around two 
hours later. Many small magnetic features with positive and negative
polarities emerge between the two main pores. Some of them cancel 
each other, but others move until they merge with the main pore of 
the same polarity. The growth rate of the two polarities is roughly 
the same, with the exception of the observing time with GREGOR, 
where the negative polarity (blue) remains stationary over 
approximately two hours. The polarities during its lifetime
were not balanced, i.e., the leading positive polarity (red) was 
dominant. This may reflect that the chosen ROI does not completely 
encompass the full bipolar region, that the flux moves across the 
boundary of the ROI, or that HMI is missing part of the flux 
(e.g., due to small inconsistencies in calibration of plage and pore fields).

Approximately one day after emergence, at 20:00~UT on 2015 April~17, 
the pores reached their maximum size and magnetic flux. While the negative 
polarity started to decay immediately, the positive polarity
reached another maximum around 16 hours after the first one. The decay 
rate after the second maximum is clearly stronger for the positive 
polarity. The leading pore with positive polarity appears more 
fragmented, the negative pore is larger and has a diameter of 
around 5\arcsec. The negative flux reaches values up to 
$3.8 \times 10^{20}$~Mx and the positive polarity
up to $5 \times 10^{20}$~Mx (see Fig.~\ref{FIG05_4}).

\begin{figure}
\includegraphics[width=\columnwidth]{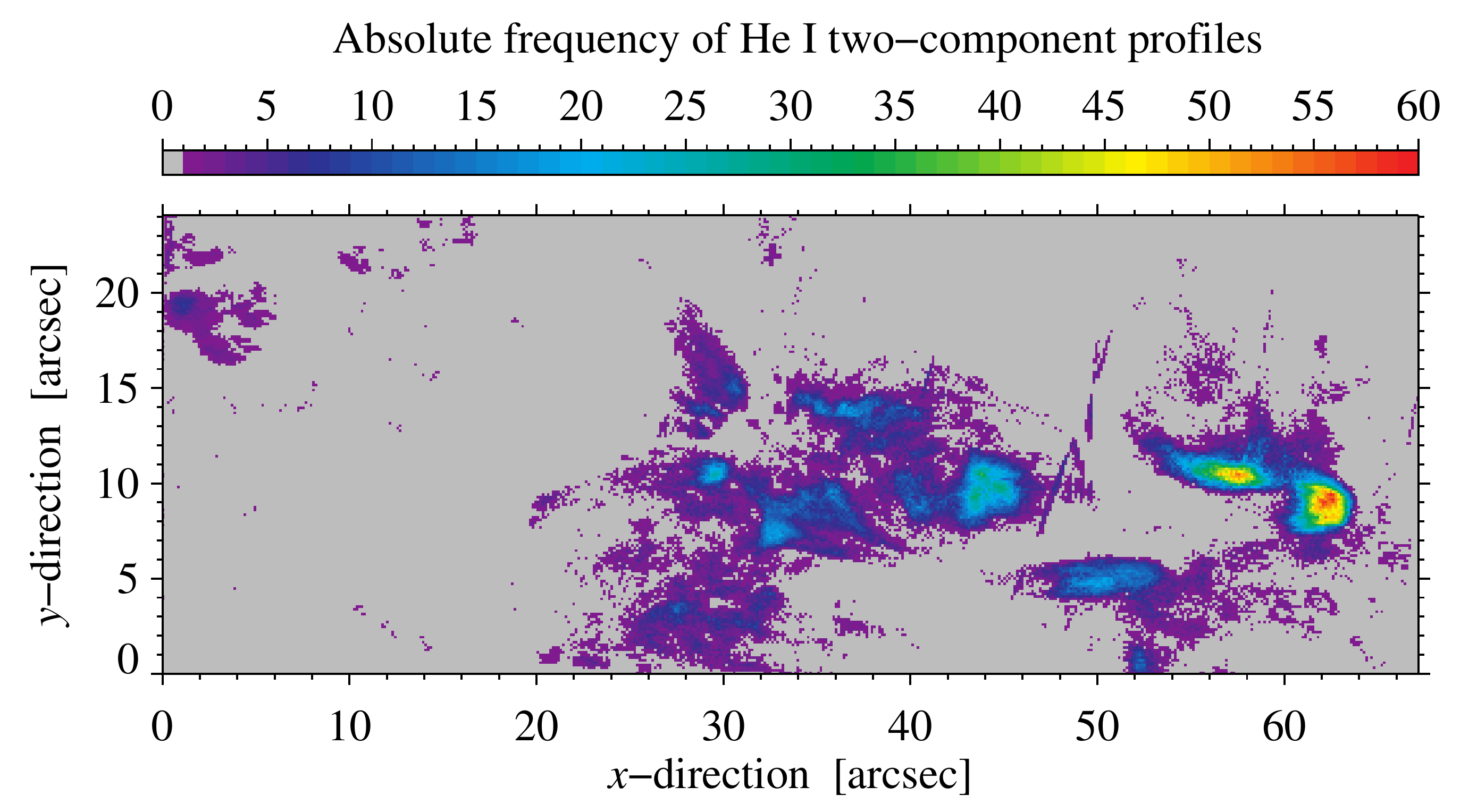}
\caption{Frequency of occurrence of dual-flow profiles during the observing 
    period. The color bar represents the number of maps in which dual 
    flows are present at a particular location.}
\label{VEL_CA_SI}
\end{figure}


\subsection{Description of the arch filament system}\label{SEC5_4_2}

An AFS is visible in the chromosphere, as is seen in the line core intensity
map of the red \ion{He}{i} spectral line (Fig.~\ref{FIG02_4}). The 
qualitative differences between line core intensity and equivalent width
(Fig.~\ref{FIG02_4}) are mainly produced by velocity broadening (or 
velocity induced splitting of the line profile into two components) 
and only to a very small degree by saturation (because the line is 
relatively weak). The arch filaments connect the two pores and other
small-scale magnetic elements that are emerging mainly between them. 
The individual arch filaments change with time during the observations on
timescales of minutes (see Sect.~\ref{SEC5_4_4}).

We investigated whether the observed AFS is visible in the corona, i.e., 
if the observed loop structures match those visible in AIA 304~\AA\ \ion{He}{ii} 
maps. Even though the \ion{He}{i} triplet (GRIS) and the \ion{He}{ii} maps 
(AIA) sample different temperatures, they may populate the same loops. We 
aligned the 64 GRIS \ion{He}{i} line depression maps with the \ion{He}{ii} 
maps closest in time. The \ion{He}{i} line depression maps are calculated 
by subtracting the average local continuum (which is constant) from the
minimum value of the red component of each \ion{He}{i} profile. This
facilitates the comparison of loops in the two maps. Pearson's linear 
correlation coefficient was computed between matching \ion{He}{i} and
\ion{He}{ii} maps. Missing points in the \ion{He}{i} maps caused by image
rotation were excluded when computing the correlation coefficients. 
The mean correlation for the 64 maps is 72\%. This 
is a significant correlation but also indicates that a one-to-one
correspondence does not exist between the loops observed 
in \ion{He}{i} and \ion{He}{ii}. Nevertheless, some loops seen 
in the neutral and ionized helium lines are likely the same loops but 
observed at different heights.

\begin{figure*}
\centering
\includegraphics[width=0.8\textwidth]{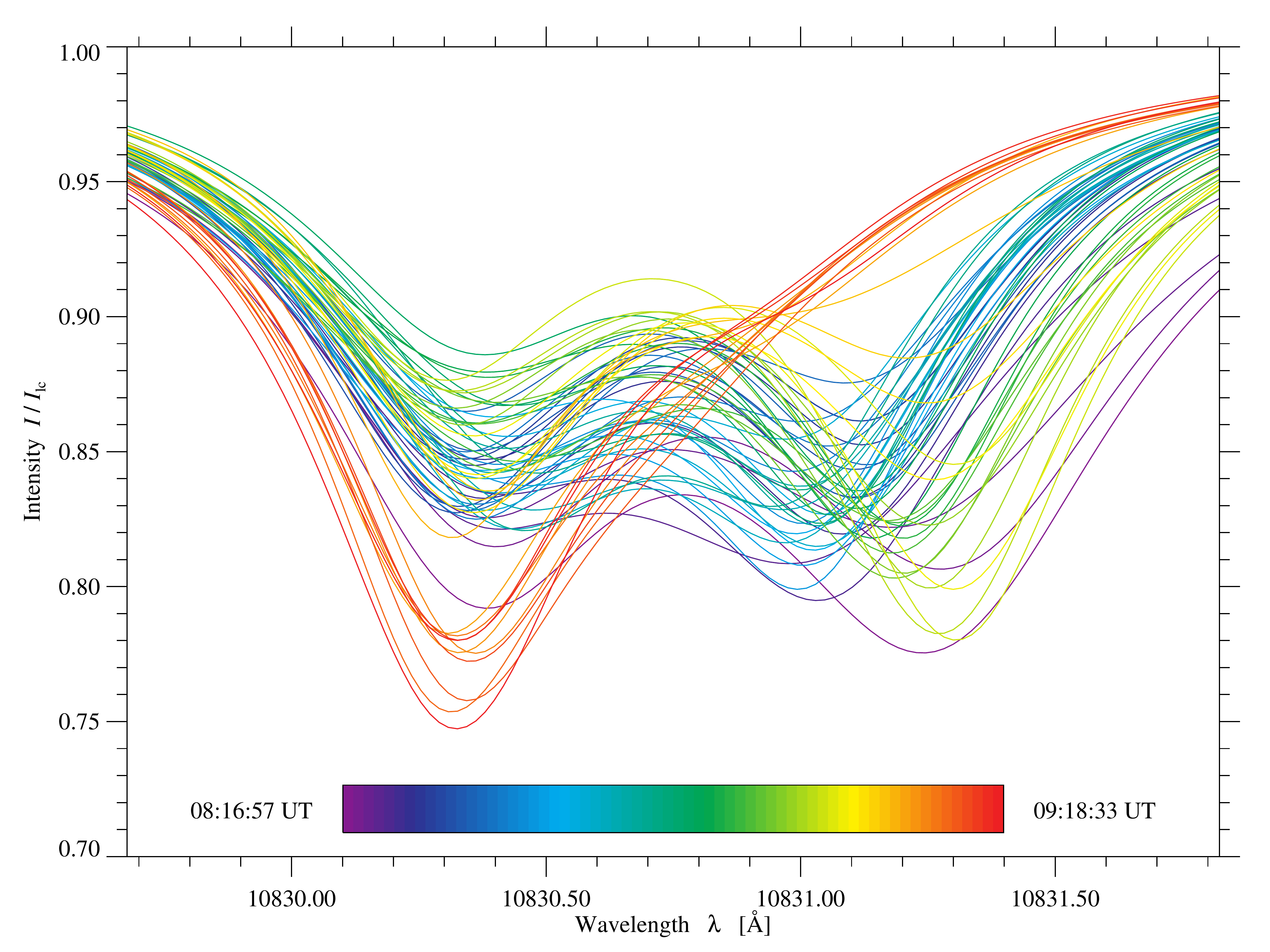}
\caption{Temporal evolution of the fitted spectral profiles (normalized to 
         the local continuum intensity $I_\mathrm{lc}$) for the red
         component of the He\,\textsc{i} triplet. The rainbow-colored
         bar indicates the elapsed time after the start of the one-hour
         time series. The spectra were taken at a strong downflow kernel
         marked by a blue filled circle in the second panel of Fig.~\ref{FIG02_4}.
         These profiles are an average of all fitted spectral
         profiles within the blue patch and do not refer to individual fits.
         The spectra were slightly smoothed to avoid a cluttered
         display.}
\label{HE_PROFILES}
\end{figure*}

\begin{figure}[t]
\centering
\includegraphics[width=\columnwidth]{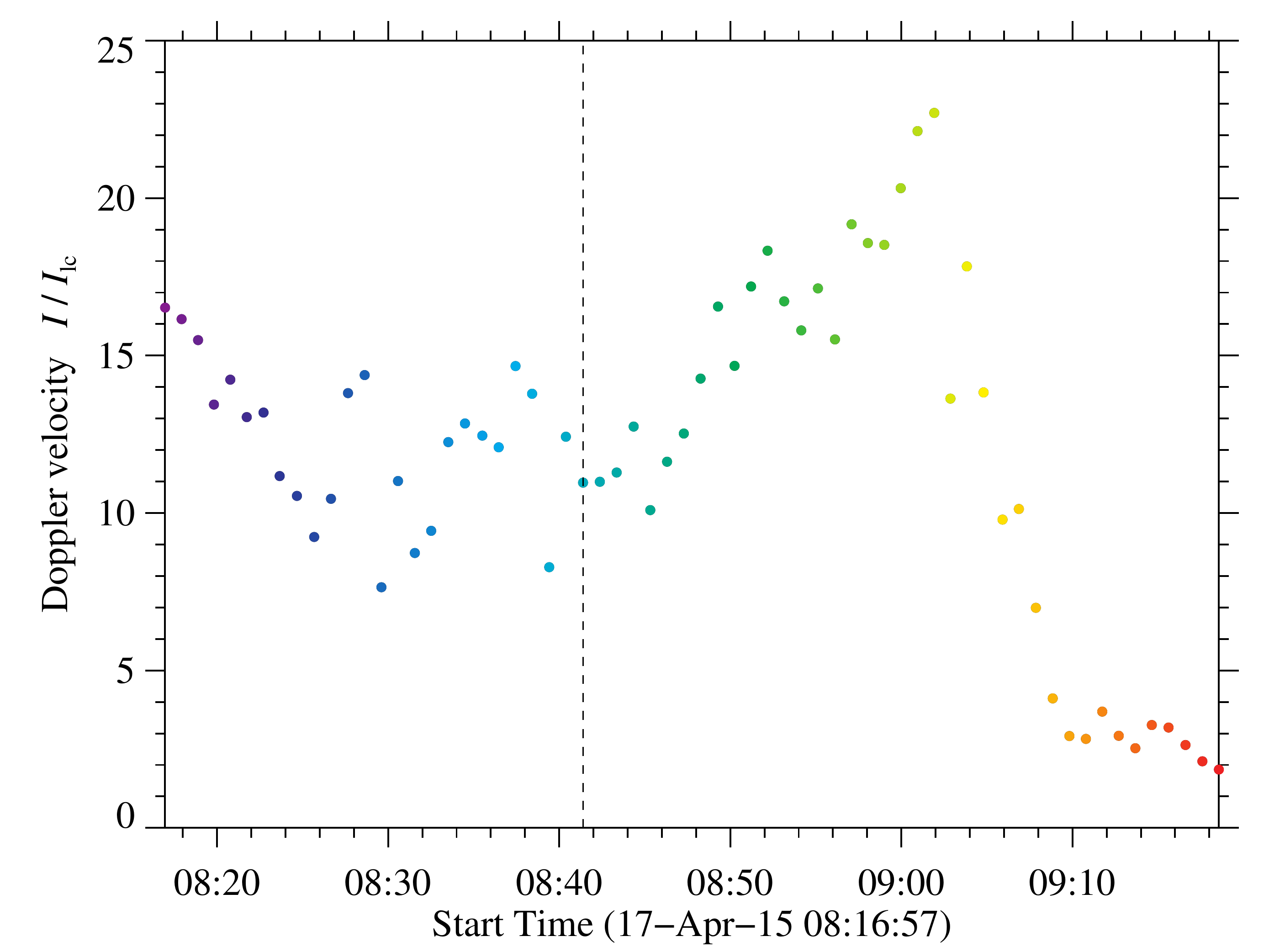}
\caption{Temporal evolution of the average Doppler velocities based on 
    the \ion{He}{i} triplet's red component in Fig.~\ref{HE_PROFILES}.
    The Doppler velocities were calculated either for a single flow
    component or refer to the fast component when dual flow components are present.
    The colors refer to the same profiles and times as in
    Fig.~\ref{HE_PROFILES}. The dashed vertical line marks the time 
    when the profile shape starts to change rapidly with time in Fig.~\ref{HE_PROFILES}.}
\label{HE_PROFILES_VEL}
\end{figure}

\begin{figure}[t]
\centering
\includegraphics[width=\columnwidth]{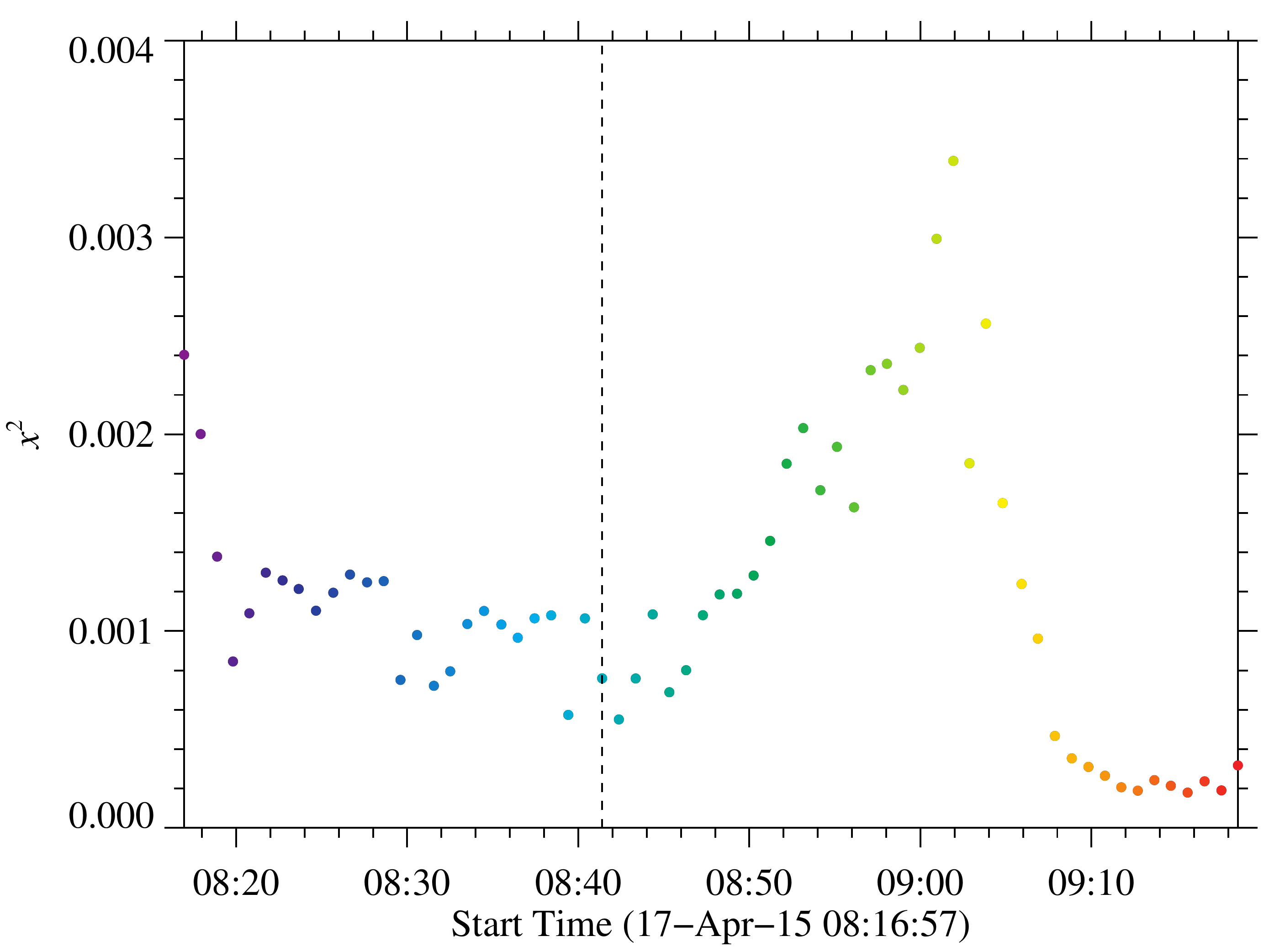}
\caption{Temporal evolution of the ${\chi}^2$-statistics for the 
          single-Lorentzian fitting method.}
\label{CHI}
\end{figure}

\begin{figure}[t]
\centering
\includegraphics[width=\columnwidth]{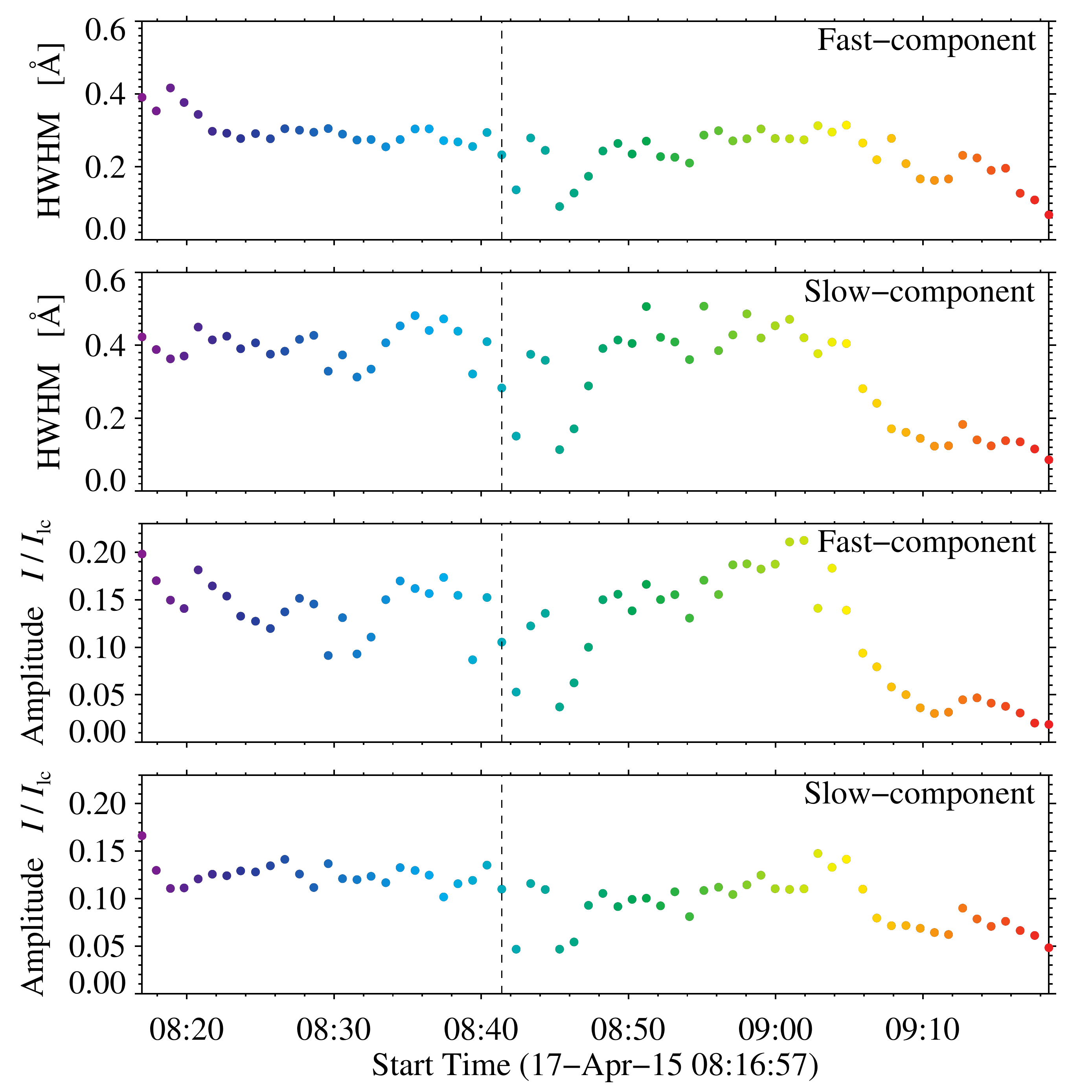}
\caption{Temporal evolution of the average HWHM of the \ion{He}{i} (red)
    fast and slow component, and the average amplitude of the fast and
    slow component measured with the double-Lorentzian method. The number of
    dual-flow profiles contained within the blue circle in Fig.~\ref{FIG02_4}
    varies with time between 18 and 69.}
\label{ALL_COEF}
\end{figure}

\begin{figure}[t]
\includegraphics[width=\columnwidth]{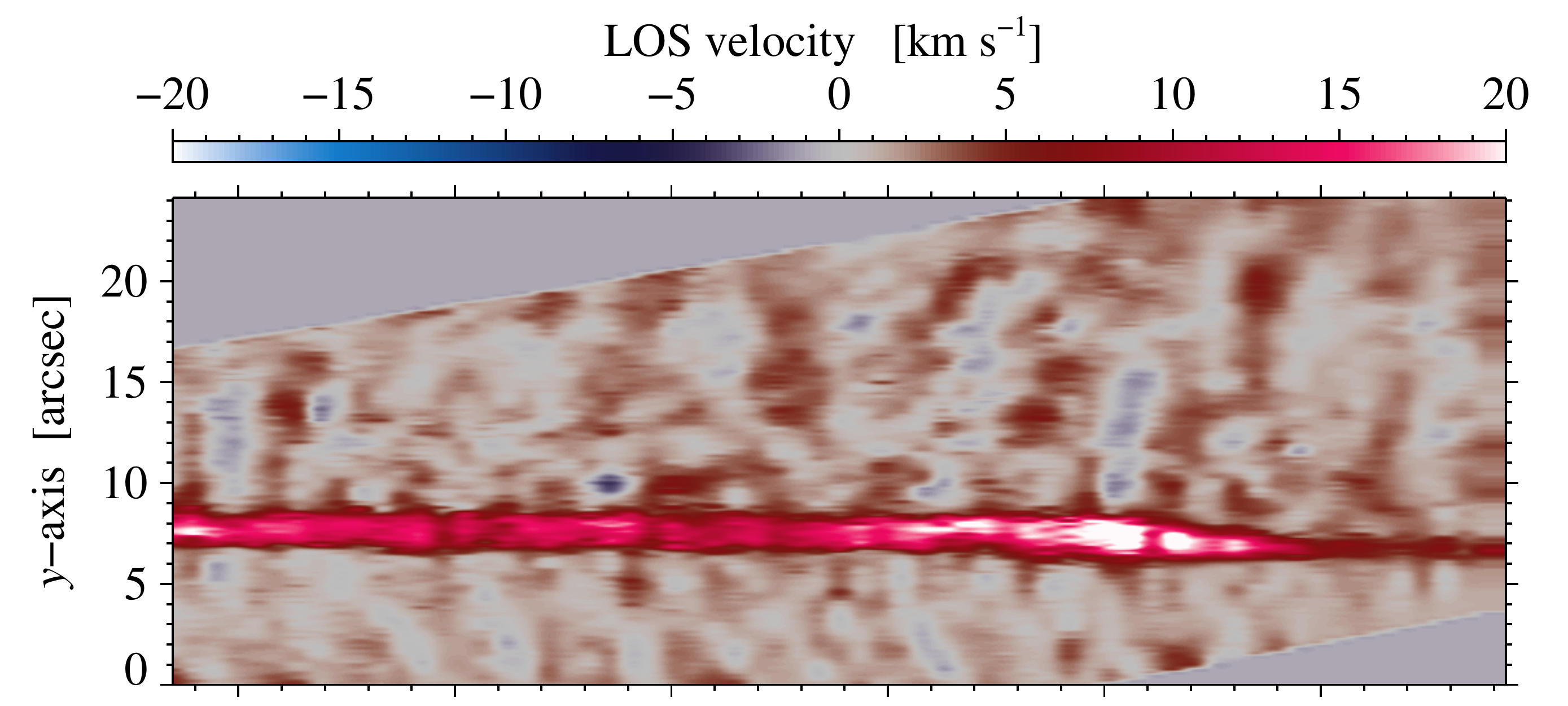}
\hfill
\includegraphics[width=\columnwidth]{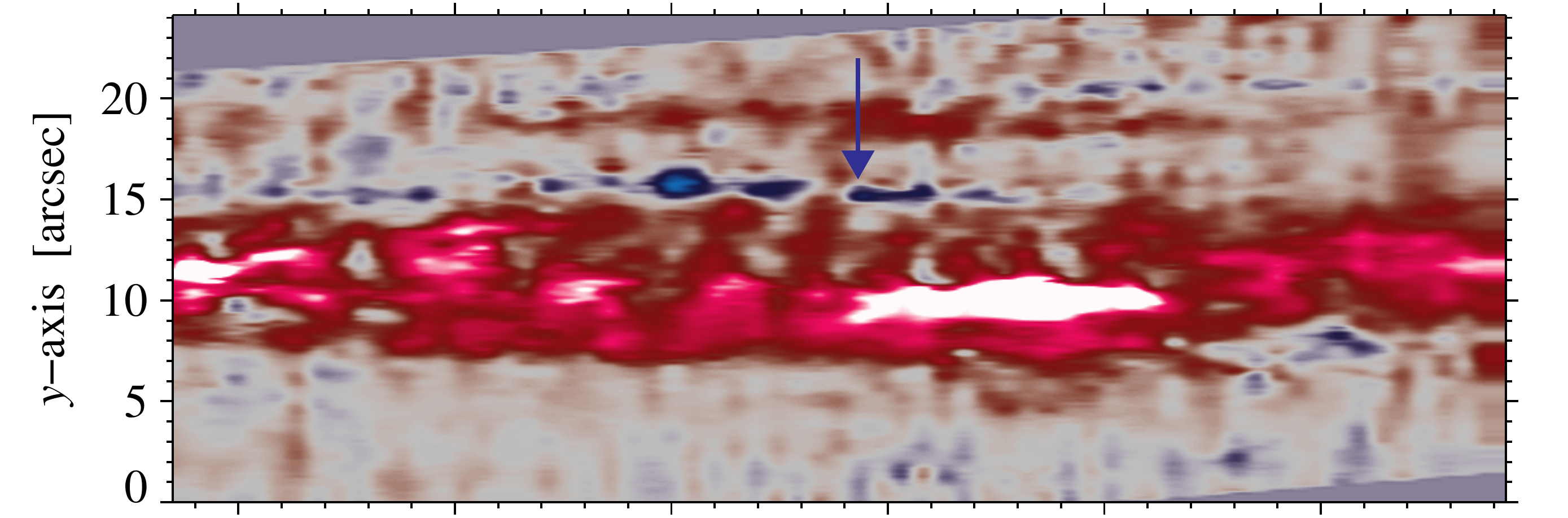}
\includegraphics[width=\columnwidth]{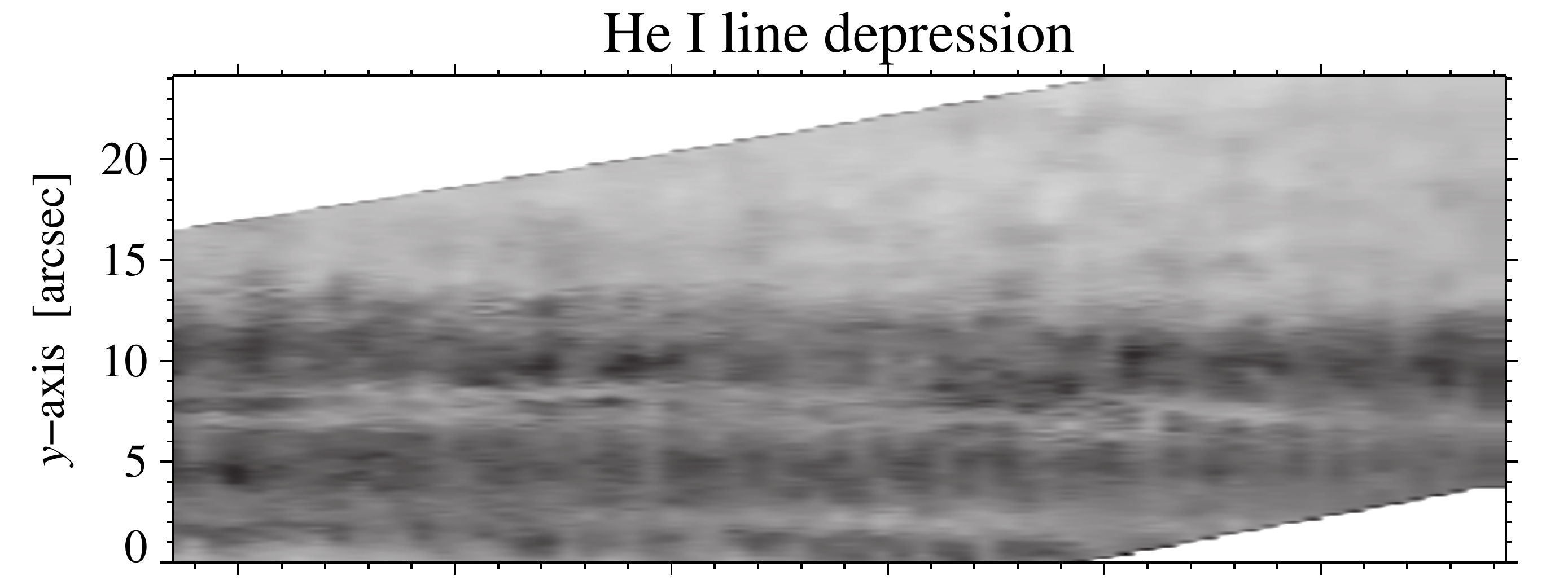}
\includegraphics[width=\columnwidth]{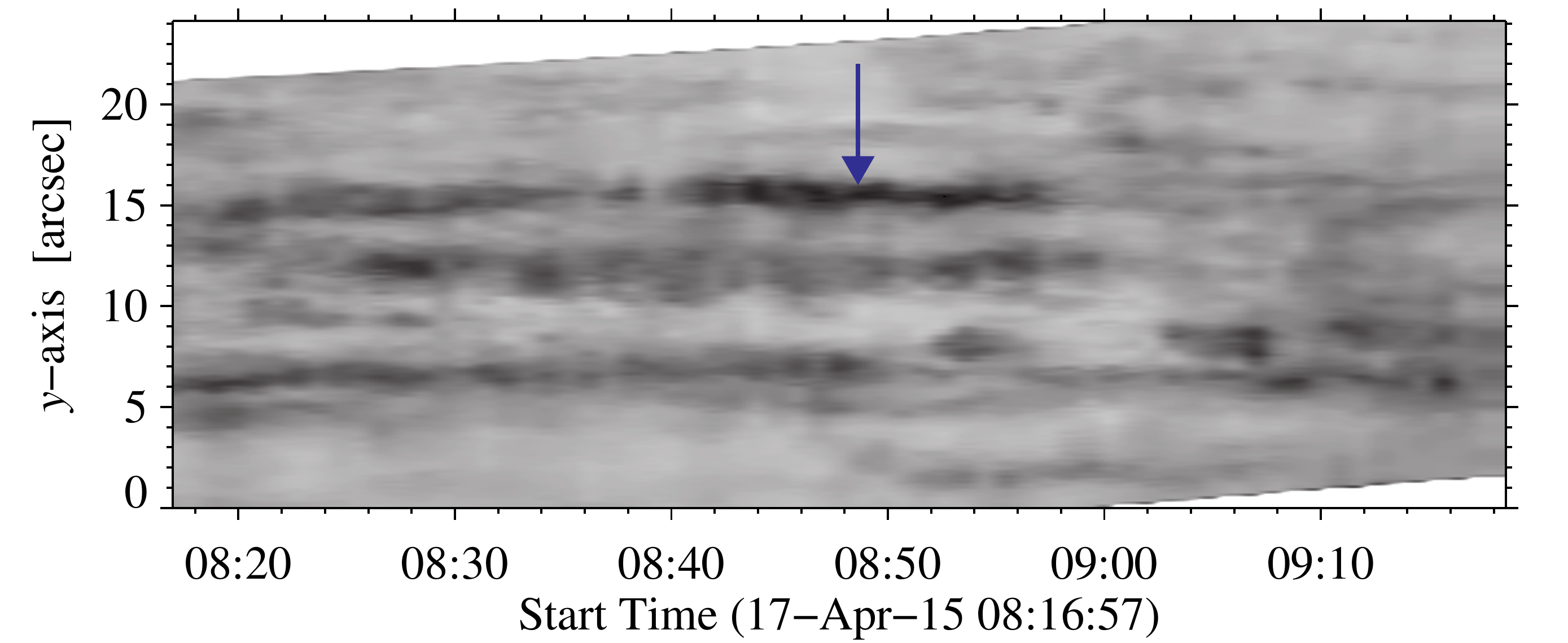}
\caption{Space-time diagrams of the He\,\textsc{i} Doppler velocity 
     based on single-component fits (\textit{top 
     two panels}) and the He\,\textsc{i} line depression (\textit{bottom two
     panels}). Both quantities are displayed for a footpoint (\textit{top and
     third panel}) and a loop top (\textit{second and bottom panel}) position
     of an arch filament. These positions are marked in Fig.~\ref{FIG02_4} by 
     vertical red lines. The reference rotation image is the same as presented
     in Fig.~\ref{FIG02_4}. The blue arrow points out the individual arch 
     filament in Figs.~\ref{FIG10_4} and \ref{FIG11}. The Doppler velocities 
     were calculated assuming single- or dual-flow components when present.
     }
\label{FIG_EVOL_INT}
\end{figure}

The time series of GRIS data covers about one hour. 
Averaging various quantities over this time interval reveals persistent 
chromospheric downflows (see movie attached to Fig.~\ref{FIG02_4}). 
In Sect.~\ref{SEC4_4.3}, we described a method to automatically identify 
dual-flow profiles in the \ion{He}{i} triplet. Their presence is indicated by contours
in Fig.~\ref{FIG02_4}. For each data set of the time series we created a 
binary mask containing the locations of dual-flow profiles. In every binary mask
the pixels with dual flow profiles are marked by one and the rest with zero.
Every map was added to obtain a final map with their frequency of occurrence shown in
Fig.~\ref{VEL_CA_SI}. Gray areas refer to regions 
where dual-flow profiles are absent, and red colors indicate that in more
than 60 maps the dual-flow profiles are present at the same location
throughout  the whole time series. Dual-flow profiles are mainly
associated with the AFS, 
and are most common near the footpoints of arch filaments. However,
at the location of pore P$_1$ they are almost absent during our observing time, 
while at the coordinates (44\arcsec, 10\arcsec) they are encountered more 
frequently for the negative polarity, and they are almost always present in the 
proximity of pore P$_2$, which exhibits the strongest downflows. One of the regions with 
dual flow profiles in Fig.~\ref{VEL_CA_SI} is circular in shape (63\arcsec, 9\arcsec), 
whereas the other  is elongated and extends along a distinct and 
persistent arch filament with strong \ion{He}{i} absorption (58\arcsec, 11\arcsec).


\subsection{Chromospheric dynamics near the leading pore}\label{SEC5_4_3}

The temporal evolution of He\,\textsc{i} 10830~\AA\ supersonic downflow 
intensity profiles near the leading pore P$_2$ was previously analyzed 
by \citet[][]{GonzalezManrique2017}, and are shown in their Fig.~4. The 
profiles correspond to an average of 69 spectra which are located inside
the blue circle with a diameter of 1.4\arcsec\ shown in the line core 
intensity \ion{He}{i} triplet map in Fig.~\ref{FIG02_4}.  In total there 
are 64 such averaged profiles, one for each slit-reconstructed image. 
In the present study, we fitted these spectral profiles using the method
described by \citet{GonzalezManrique2016} and explained in
Sect.~\ref{SEC4_4.3}.
Accordingly, we fitted the 69 individual spectral profiles in every map,
either with single- or dual-flow component Lorentzian fits, whichever is 
more appropriate. The number of dual-flow profiles within the blue circle 
varies between 18 and 69 during the time period from 08:16:57~UT to
09:18:33~UT. Thus, the average Doppler velocities contains both single- 
and double-peaked red components of the \ion{He}{i} red profiles. In
Fig.~\ref{HE_PROFILES}, we show the temporal evolution of the fitted 
spectral profiles. The inferred Doppler velocities are
shown in Fig.~\ref{HE_PROFILES_VEL}, where every point in
Fig.~\ref{HE_PROFILES_VEL} corresponds to the average velocity of 
the 69 pixels within the blue circle in Fig.~\ref{FIG02_4}.

At the beginning of the time series, the slow and fast components had their
minima well separated, forming a \textsf{w}-shaped profile with similar
intensities in the line core (see Fig.~\ref{HE_PROFILES}). In the 
first 25 minutes of the times series the average values of the 
Doppler velocities fluctuated in the range of 7\,--\,17~km~s$^{-1}$.
The shape of the profiles changed very quickly with time. After 25 minutes, near
08:41~UT (marked as dashed line in Fig.~\ref{HE_PROFILES_VEL}), 
the minima of the two components are no longer 
well separated and the intensity profile of the fast component becomes deeper. 
Starting around 08:41~UT  the mean
Doppler velocities also increased, reaching a maximum of about 23~km~s$^{-1}$.
The difference between line-core intensities of the two components reaches 
$\Delta I / I_0 \approx 0.1$. At the end of the time series, near 09:08~UT, 
the profiles change suddenly into single-component profiles with deeper
line-core intensities than before. At this time, the majority of the
profiles within the  blue circle are single-component profiles.  As a
consequence, the Doppler velocities drop drastically to 2\,--\,4~km~s$^{-1}$.
This is consistent with the temporal evolution of the ${\chi}^2$-statistic for
the single-Lorentzian fitting method (see Fig.~\ref{CHI}) used to fit all 69
\ion{He}{i} red component profiles within the blue circle in
Fig.~\ref{FIG02_4}. The average ${\chi}^2$ values are lower at the end of the
time series, when more single-flow profiles are present.

The evolution of the half width at half minimum (HWHM) and
the amplitude for the fast and slow components of the \ion{He}{i} (red)
component is measured with a double-Lorentzian 
\citep[see][Sect.~3, for more information]{GonzalezManrique2016}. 
The HWHM for the slow and fast components exhibit a similar behavior. The
width of the slow component usually exceeds the value of the fast component.
Minimum values appear at the end of the time series and for a few minutes
after 08:40~UT. At this time, the number of dual-flow profiles dropped to
18\,--\,22. An analogous behavior is observed for the amplitude of both the
slow and fast components. The highest values correspond to profiles in
Fig.~\ref{HE_PROFILES} around 09:05~UT. The lowest parameters for the two
components both match in time with the minimum values of the HWHM.


\subsection{Tracking individual arch filaments}\label{SEC5_4_4}

The second and bottom panels  of Fig.~\ref{FIG_EVOL_INT} displays the temporal
evolution close to the loop top of an arch filament, whereas the top and 
third panels show the temporal evolution of a footpoint close to the pore
P$_2$.
The positions are marked by the two red
vertical lines in Fig.~\ref{FIG02_4}. The darkest region in the lower panel of
Fig.~\ref{FIG_EVOL_INT} matches in time with the highest 
upflows in the second panel. An extended discussion of this figure is
presented in Sect.~\ref{SEC6_2}.

\begin{figure}[t]
\includegraphics[width=\columnwidth]{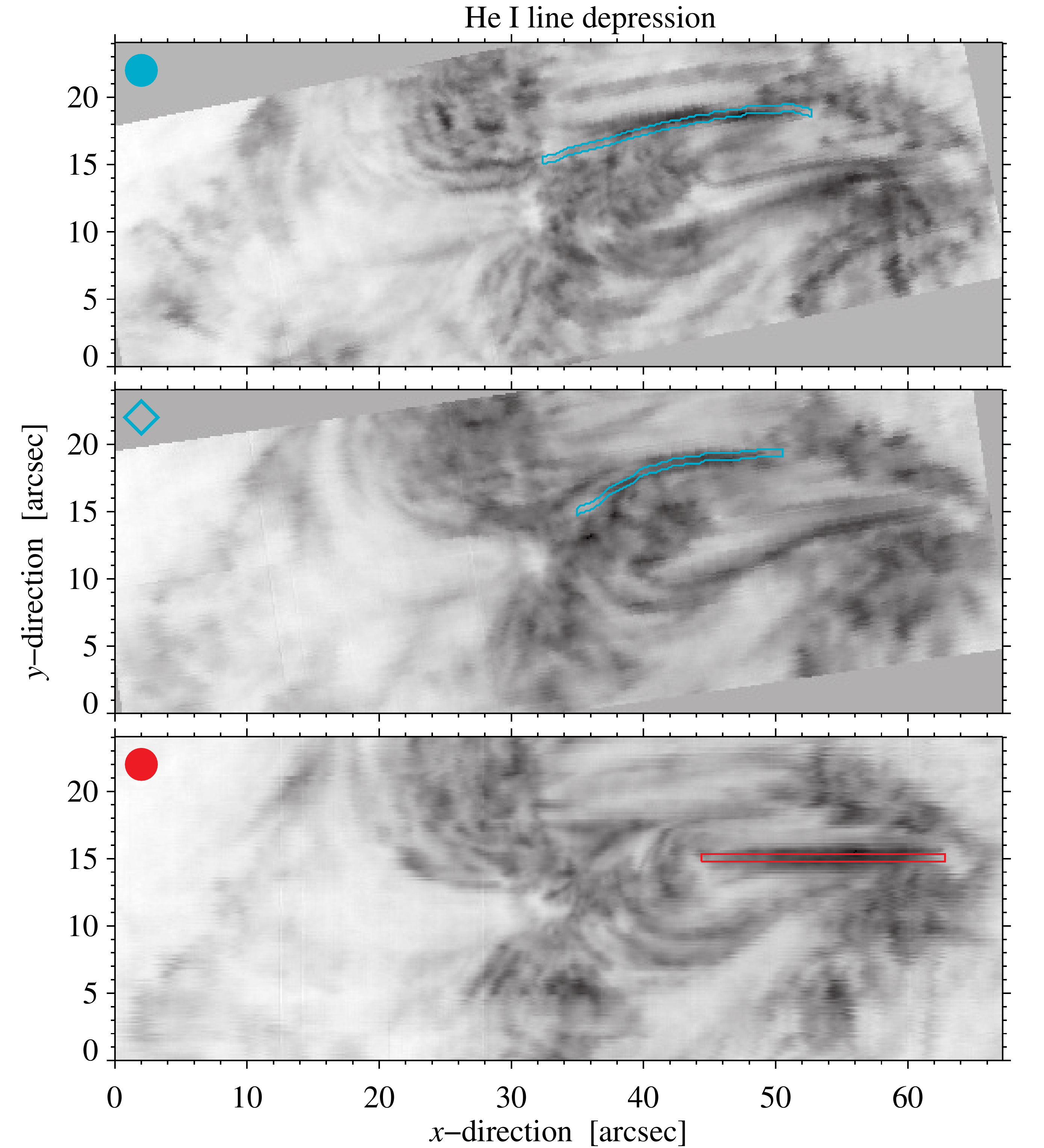}
\caption{\ion{He}{i} line depression map at 08:16:57~UT (\textit{bottom}), 
    08:40:23~UT (\textit{middle}), and 08:49:16~UT (\textit{top}) on 2015 
    April~17. The contours outline different arch filaments at different
    times. These images are rotated with respect to the reference image
    at 08:16:57~UT. The symbols and colors correspond to the arch filaments in 
    Fig.~\ref{FIG09_4}.}
\label{FIG08_4}
\end{figure}

\begin{figure}[t]
\includegraphics[width=\columnwidth]{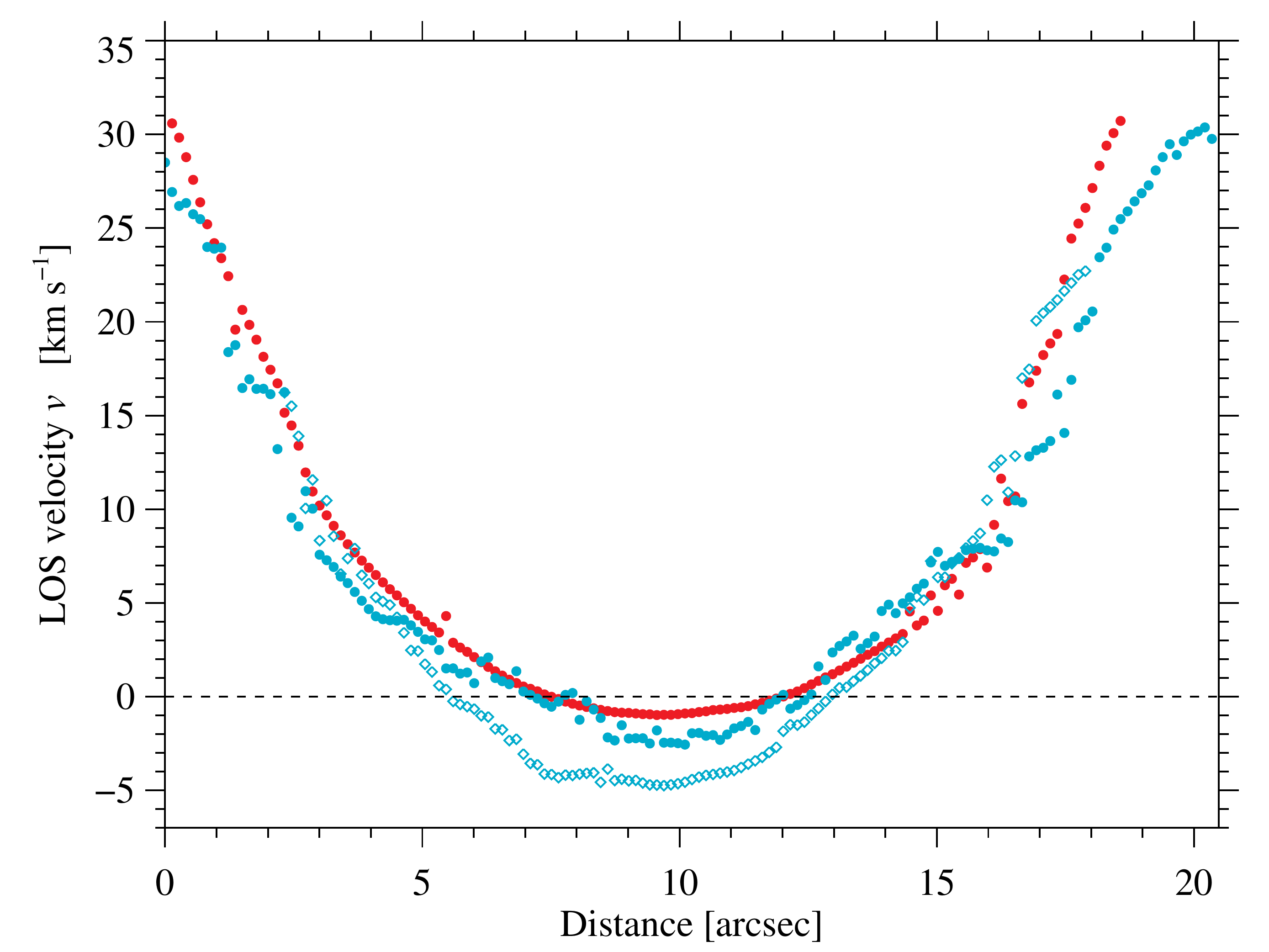}
\caption{Doppler velocities inferred from the \ion{He}{i} triplet's red 
   component. The Doppler velocities were calculated assuming single- or 
   dual-flow components, whichever is more appropriate.
   The velocities are calculated along three arch filaments shown in
   Fig.~\ref{FIG08_4}. The colors of the contours and symbols 
   (\textit{upper left corners of panels in Fig.~\ref{FIG08_4}}) 
   correspond to the contour lines in that figure.
   }
\label{FIG09_4}
\end{figure}

We tracked in time the average Doppler velocities along an individual 
arch filament. Figure~\ref{FIG08_4} contains three \ion{He}{i} line depression
maps at three different times. The contours depict three arch filaments at
08:16:57~UT (red contour/bullet), 08:40:23~UT
(blue contour/square), and 08:49:16~UT (blue contour/bullet) on 2015
April~17. The two blue contours represent the same arch
filament at different times. We inferred the Doppler velocities along the
contours and calculated the mean values of five pixels in the $y$-direction
for every position in the $x$-direction. The mean LOS
velocities for all three arch filaments are shown in 
Fig.~\ref{FIG09_4}. The loop tops show upflows (negative velocities) up to
6~km~s$^{-1}$  and 2~km~s$^{-1}$ in the case of the blue and red arch
filaments, respectively. The distance between opposite-polarity 
footpoints  are up to 20\arcsec\ for the two arch filaments.

\begin{figure}[t]
\includegraphics[width=\columnwidth]{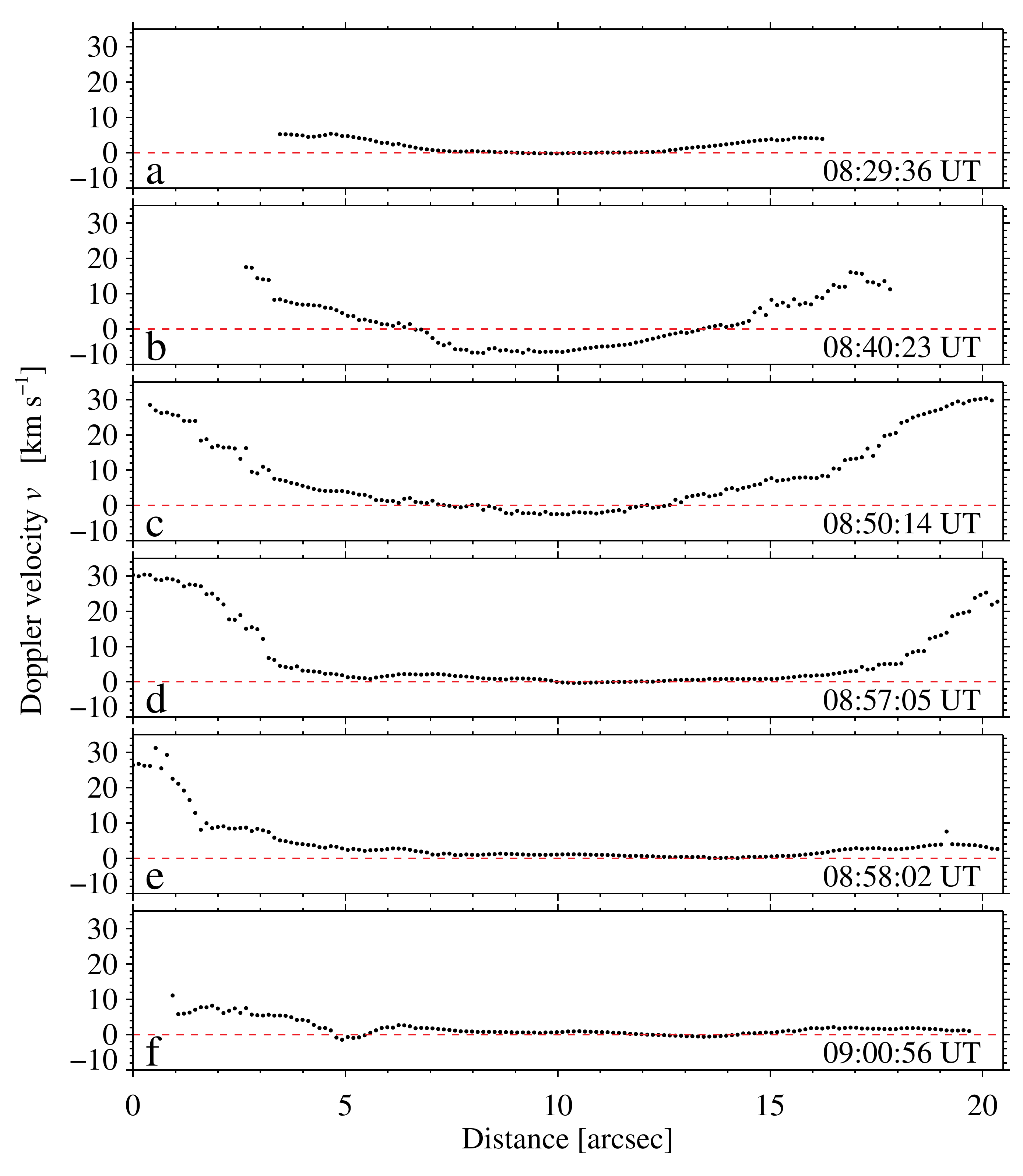}
\caption{Temporal evolution of the Doppler velocities along an individual arch
    filament (blue contours in Fig.~\ref{FIG08_4}) based on the \ion{He}{i}
    triplet's red component. The Doppler velocities were calculated 
    based on either single- or dual-flow components.}
\label{FIG10_4}
\end{figure}

Next, we follow the temporal evolution of one individual arch 
filament and calculate the average Doppler velocities along
it. We selected the arch filament outlined by a blue contour in
Fig.~\ref{FIG08_4}.
This arch filament is special since it emerged and disappeared
during our observing sequence. Figure~\ref{FIG10_4} shows
the temporal evolution of the average Doppler velocities along 
the blue arch filament at six different moments. The lifetime 
of the arch filament is about 25\,--\,30~min, which is at the upper end of
values (10\,--\,30~min) given in the literature \citep[e.g.,][]{Chou1993}.
Initially, the loop tops are nearly at rest, while there are small downflows 
near the footpoints. After approximately 10~min the loop tops 
exhibit the fastest average upflows (in the range of 5\,--\,6~km~s$^{-1}$) 
and supersonic downflows with mean values of up to 31~km~s$^{-1}$ are present
at the footpoints. Near the end of the time series, the velocities
approach zero in the whole arch filament. The upflows stop first, while the 
downflows still persist for another 20~min. 
In addition, the distance
between the footpoints increases with time. At the beginning, the 
length of the arch filament was roughly 13\arcsec\ while at the end of its
lifetime the distance amounts to 21\arcsec.

We followed the temporal evolution of the average Doppler velocities 
at the loop tops and footpoints. The results are represented 
in Fig.~\ref{FIG11} (\textit{lower panel}). 
The black bullets represent the mean Doppler velocities
at the loop top while the blue and red bullets indicate the 
mean Doppler velocities at the left and right
footpoints, respectively. To calculate the average of the 
Doppler  velocities at the loop tops, we selected
the lowest negative value at the loop tops and the velocities 
in the two neighboring pixels on each side with a total of 
five values in Fig.~\ref{FIG10_4}. For the two footpoints, we  
selected five values, with the exception
of the first and last pixels of the blue arch filament in Fig.~\ref{FIG10_4}. 
The individual arch filament appears in 33 out of 64 maps.

In the middle panel of Fig.~\ref{FIG11}, we also followed 
the temporal evolution of the length of the
individual arch filament during the same period. The evolution of the
average \ion{He}{i} line depression along the arch filament is also
represented in the upper panel of Fig.~\ref{FIG11}. The values were 
normalized to one. A value of unity indicates the darkest part of the
filament.

We found that the strongest upflows at the loop tops started at about 8:40~UT, 
and one minute later the downflows at the footpoints
started to increase. Shortly before 08:40~UT the value of \ion{He}{i} 
line depression was the lowest. After this time, the values increased. 
The length of the arch filament also increased at the same time. 
These downflows became very strong 
(more than 20~km~s$^{-1}$ on average) 
after two minutes in the case of the left footpoint (blue bullets in
Fig.~\ref{FIG11}) and after around four minutes in the case of the right
footpoint (red bullets in Fig.~\ref{FIG11}). 
The highest value of the \ion{He}{i} 
line depression was also after four minutes. Afterwards the values slowly
decreased.
The downflows reached their highest velocity at the footpoints between
10\,--\,15~min after the upflow was detected at the loop tops. The arch
filament was longest also about 15~min after the upflow was detected
(08:55~UT). Afterwards, the downflows started to decay until the 
arch filament fully disappeared at 09:00~UT, i.e., 20~min after the downflows
with the highest velocity were detected.

%
%

\section{Discussion}\label{SEC6}


\subsection{Supersonic downflows near the leading pore}\label{SEC6_1}

In the following, we investigate the temporal evolution of a 
chromospheric AFS connecting two photospheric pores. 
HMI magnetograms revealed both flux emergence
and decay during the lifetime of the pores, which lasted about 
three days. We followed the dynamics of the AFS, where the 
Doppler velocity reached up to to 40~km~s$^{-1}$ in single pixels.
At the end of the observing period, the average 
values of the Doppler velocities dropped to 2\,--\,4~km~s$^{-1}$. 
Several studies of AFS also based on the \ion{He}{i} triplet reported
supersonic downflows
\citep[e.g.,][]{Schmidt2000b, Solanki2003b, AznarCuadrado2005,  AznarCuadrado2007, Sasso2007, Lagg2007, Balthasar2016, GonzalezManrique2016, GonzalezManrique2017}.

\begin{figure}[t]
\includegraphics[width=\columnwidth]{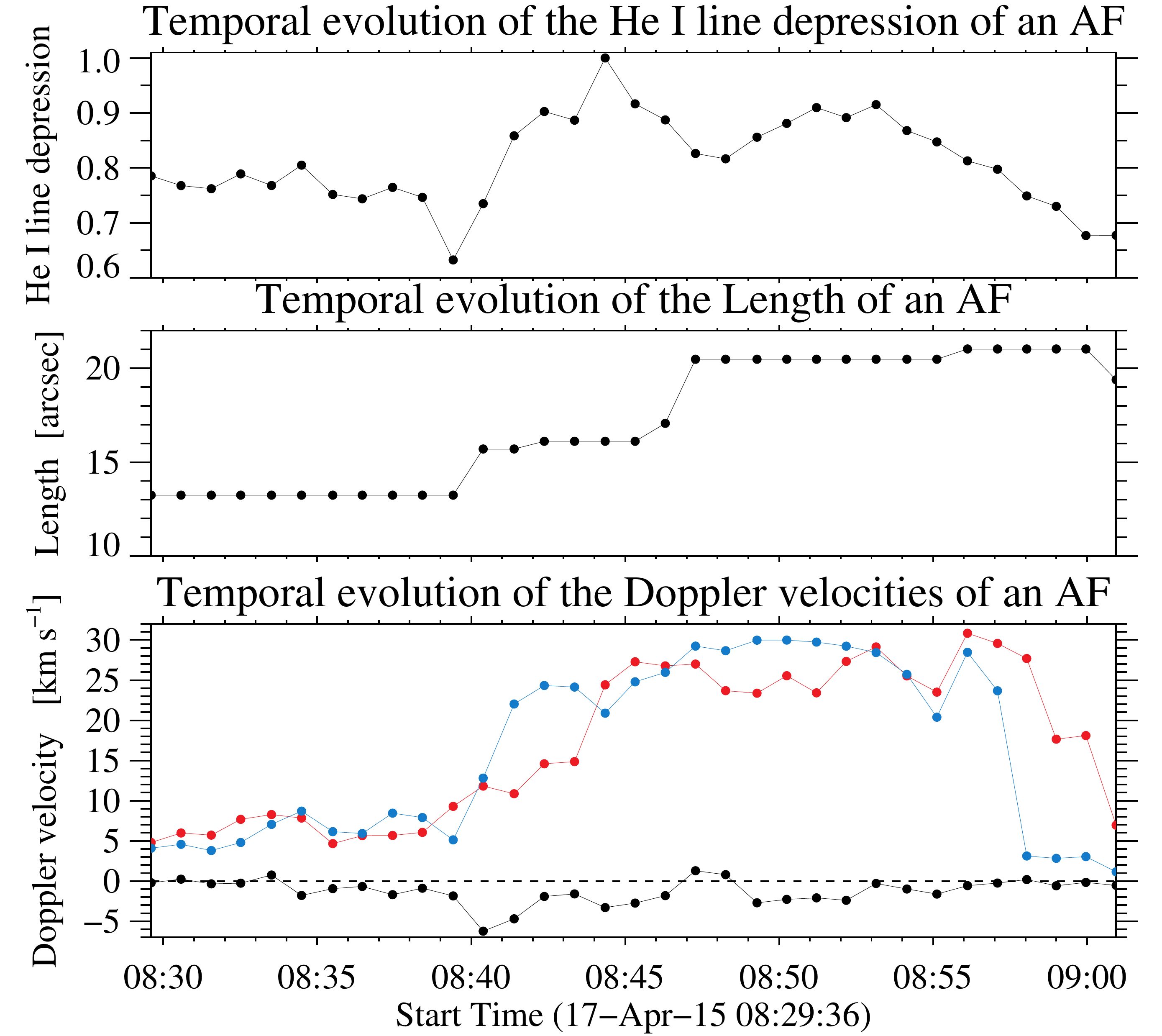}
\caption{Temporal evolution of the normalized mean \ion{He}{i} line depression
   of the whole arch filament (\textit{upper panel}). The highest value is the darkest. 
   Temporal evolution of the length of the individual arch filament (\textit{middle panel}). 
   Temporal evolution of the mean Doppler velocities of the individual arch filament 
   (\textit{lower panel}) shown in Fig.~\ref{FIG10_4}. The Doppler velocities were calculated 
   based on either single- or dual-flow components. The black bullets represent
   the mean Doppler velocities at the loop top and the blue and red bullets 
   indicate the mean Doppler velocities at the left and right footpoints, respectively.}
\label{FIG11}
\end{figure}

\citet{Lagg2007} studied a similar case with supersonic 
downflows close to a growing pore. They reported on 
supersonic downflows at the footpoints of an emerging 
magnetic loop with LOS velocities of up to 40~km~s$^{-1}$, 
as in our case. They observed during a time interval of about 70~min. 
Moreover, they studied a region near a pore with a small FOV (3 $\times$ 3~Mm)
of one-minute and five-minute cadence for a total of 18~min with all the four 
Stokes parameters. In our case, scanning the ROI results in a one-minute
cadence for the one-hour observing sequence, but without polarimetry. 
\citet{Lagg2007} observed persistent dual red components of the \ion{He}{i}
triplet. One slow component was always almost at rest while the fast 
component showed supersonic velocities within the same resolution element. 
We found exactly the same behavior near
our growing leading pore (see Fig.~\ref{HE_PROFILES}).

\begin{figure*}[t]
\centering
\includegraphics[width=\textwidth]{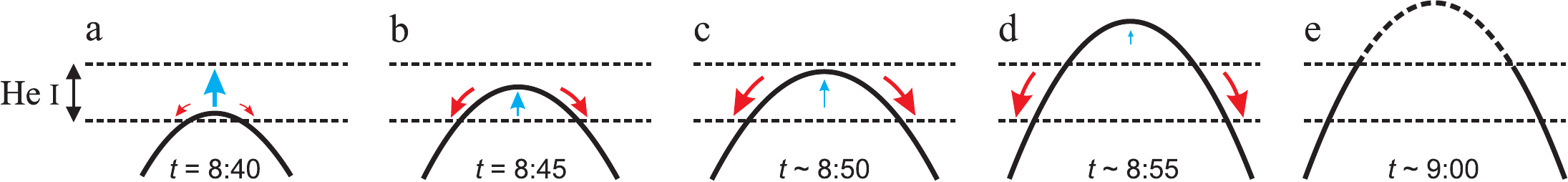}
\caption{Cartoon of the evolution of an individual arch filament (\textit{from a to e}). 
    Arrows in blue and red mark the direction of the plasma flows as observed in the
    \ion{He}{i} triplet, respectively. The stronger the downflows (\textit{red}), the thicker the
    arrow. The \ion{He}{i} layer lies between the two dashed lines.}
\label{FIG12}
\end{figure*}

\citet{Lagg2007} speculate that a flux tube is drained as a consequence of the
pressure balance between the flux tube that is rising from the photosphere and
the neighboring atmosphere. As a consequence, the hydrostatic support 
is lost and the material moves downwards to the solar surface. 
The theoretical draining time of the material for an individual arch 
filament is about 20~min \citep{Chou1993}. This time is consistent with the 
estimations of \citet{Xu2010}. \citet{Chou1993} conjectured that the 
lifetime of an individual arch filament is between 10\,--\,30~min, but the
entire AFS will be visible as long as  magnetic flux continues emerging. 
The lifetime of the arch filaments observed by us is in agreement with
\citet{Chou1993}. In the present study, the observations 
agree with the findings of \citet{Lagg2007} that the upflows are located 
at the loop tops and the downflows are along the legs of the arch 
filaments.

Why are supersonic downflows seen near the pore? 
Typically, downflow speeds are larger when the atmosphere is 
cooler (as in the pore), since the final stratification will have 
less gas at chromospheric heights (due to the smaller pressure scale height). 
In addition, if the field strength is large (as in the pore), then the feature
is strongly evacuated in the photosphere due to horizontal pressure balance. 
That means again that more gas has to be removed from the overlying arch 
filament, until hydrostatic equilibrium sets in. 


\subsection{Evolution of an individual arch filament}\label{SEC6_2}

We agree with the assumptions of \citet{Chou1993} and \citet{Lagg2007} 
who interpreted the strong downflows as a consequence of the draining of the
rising loops that show up as arch filaments. To demonstrate that this is the
most plausible hypothesis we studied the evolution of an individual arch
filament (in another region of the FOV) that clearly emerges and decays during
our GREGOR observing period.

One clear example of the emergence and decay 
of an individual arch filament is depicted with blue contours in 
Fig.~\ref{FIG08_4}. Its lifetime was  about 30~minutes.  
During the emerging process the arch filament gradually gets darker, and
while decaying it gets less dark and finally disappears. 
This behavior is clearly reproduced in the \ion{He}{i}
space-time diagram in Fig.~\ref{FIG_EVOL_INT} marked with a blue arrow.  
In particular, the second  and fourth panels (from the top) follow 
the evolution of the loop tops. 
The individual arch filament appears at 15~\arcsec\ (y-axis). 
The  arch filament  becomes brighter with time  because the material has drained,  because it has 
moved out of the height range over which \ion{He}{i} is formed, 
or because the gas has started getting hotter. Strong downflow LOS velocities 
appear at the same time as the arch filament became
darker. The downflows  disappear  at the same time as the arch filament
gets less dark and finally vanishes. The length of the arch filament 
changed during its lifetime.

We followed the evolution of the average Doppler velocities 
along the individual arch filament (Fig.~\ref{FIG10_4}) and focused 
on the loop top and the two footpoints (Fig.~\ref{FIG11}).
Moreover, we also monitored the variations in the \ion{He}{i} 
line depression and of the length of the arch filament.
First, upflows of up to 6~km~s$^{-1}$ were detected at the loop tops.
This value is comparable to the upflows found for example by
\citet{Lagg2007} and \citet{Xu2010}, but is smaller than the LOS velocity of
12~km~s$^{-1}$ found by \citet{Mein2000} from the IR \ion{Ca}{ii} 8542
\AA. The \ion{Ca}{ii} spectral line is formed lower in the atmosphere
than the \ion{He}{i} triplet. Additionally, the high velocity values in
\ion{Ca}{ii} can be attributed to AFSs behaving differently in different
spectral lines. It is also possible that different observed AFSs behave 
differently. 

Around 2\,--\,4~min later,
strong downflows appeared at the footpoints which quickly increased to 
supersonic speeds, on average, up to 30~km~s$^{-1}$. Simultaneously, the
\ion{He}{i} line depression is darkest. Similar results regarding the LOS
velocities were presented by, e.g., \citet{Solanki2003b}, \citet{Lagg2007}, 
and \citet{Xu2010}. The upflows significantly decreased two minutes after 
they reached their maximum value. Conversely, the downflows reach their
maximum values at the footpoints about 10--15~min later. The filament is
longest  about 15--20~min after the strong downflows appeared. The arch
filament completely vanished 20~min after the strong upflows were measured
(08:40~UT). The observations match the interpretation by 
\citet{Chou1993} who conjectured  that without upward transportation of
material the loop could be emptied by drainage in around 20~min. Our
observations also fit with the theoretical estimations made by 
\citet{Xu2010}, who estimated that the loop is emptied by
drainage in 20~min.

The observational evidence presented in this work leads to the scenario
presented in the cartoon shown in Fig.~\ref{FIG12}. The cartoon 
represents the evolution of an arch filament (from a to e) similar to the
individual filament studied here (in Figs.~\ref{FIG10_4} and \ref{FIG11}).
This cartoon only represents the evolution seen in the \ion{He}{i}.

In panel a, the individual arch 
filament is shown as it appears in \ion{He}{i} line depression images 
($t =$ 08:40~UT). The arch filament carries relatively cool plasma as 
it rises from the photosphere to the chromosphere. This panel represents 
the moment when the loop tops reach their maximum average LOS velocity
(upward). In panel b the arch filament continues to rise, but with 
lower LOS velocities at the loop tops ($t =$ 08:45~UT). 
The cool material starts to drain toward the photosphere 
along the legs of the arch filament. The distance between the footpoints 
slightly increases. In panel c, the arch filament continues to
rise with lower LOS velocities at the loop tops ($t =$ 08:50~UT).
At the same time, the velocities at the legs increase and supersonic flows 
(> 10~km~s$^{-1}$) are already present close to the footpoints. The arch
filament continues to expand horizontally. In panel d we speculate that 
the loop top has broken through the original \ion{He}{i} formation height
($t =$ 08:50~UT), but \ion{He}{i} is also present at the apex of the loop 
and down its sides \citep[as found by ][]{Merenda2011}. Formation 
of \ion{He}{i} along the loop is due to the higher density in the loop 
than in its surroundings at this stage. As more material drains down 
the loop legs, the density in the loops continues to decrease. Hence, 
at some point the loop stops being dark in the He line core 
and we stop seeing major downflows along the legs. The last panel shows how 
the loop tops of the arch filament are disappearing from the \ion{He}{i} 
observations. The material in the loop top is already 
in the higher layers of the atmosphere and the density must 
also have decreased there substantially. The average velocities of 
the loop tops are now very close to zero, as inferred from the 
\ion{He}{i} triplet. The downflows at the footpoints and the 
upflows at the loop tops gradually decrease until 
the material within the flux tube is empty and the arch filament vanishes 
($t = 09:00$~UT). Typically, individual arch filaments are 
replaced by new arch filaments, which emerge from below if enough 
flux is still available \citep{Chou1993}. In line with \citet{Chou1993} 
and \citet{Xu2010}, our arch filament vanishes after 20~min counting 
from the strongest upflows seen in the loop tops (Figs. \ref{FIG_EVOL_INT} 
and \ref{FIG11}).

%
%

\section{Summary and conclusions}\label{SEC7}
We observed the temporal evolution of an arch filament system embedded in
a bipolar emerging flux region with high cadence using the very fast
spectroscopic mode of the GRIS instrument attached to the GREGOR 
telescope. One scan was carried out in about one minute and consisted of 180
slit positions with a step size of 0.135\arcsec. The spectral region included
the chromospheric \ion{He}{i} 10830~\AA\ triplet. We followed the AFS, which
connected opposite polarities, during 64 minutes. The region did not carry
enough flux to fully develop an active region. 

During the observations, we detected several individual arch filaments. 
One filament was thoroughly tracked over its full lifetime of about 30 min,
from the moment it appeared until it vanished. At the beginning, we inferred
upflows in the loop top between 5 and 6~km\,s$^{-1}$. After 2--4~min,
supersonic downflows started to appear with Doppler velocities in the 
range of 20--40~km\,s$^{-1}$. The arch filament expanded with time by about
7\arcsec\ in length during its lifetime (30~min). Our observations support
predicted AFS scenarios by \citet[][and references therein]{Chou1993} who
hypothesized that the downflows in arch filaments could be explained by the
emergence into the solar atmosphere of a flux tube. \citet{Lagg2007} 
suggested that the high-density gas carried by the rising loop drains 
down its legs due to the action of gravity and the simultaneous need for
horizontal pressure balance and vertical hydrostatic 
equilibrium. As shown in the cartoon (Fig.~\ref{FIG12}), we follow 
the arch filament as it carries plasma during its rise from the photosphere 
to the corona. The material then drains toward the photosphere, reaching
supersonic velocities, along the legs of the arch filament.

Near the leading pore (P$_2$), persistent chromospheric supersonic downflows
were found for about 60~min. The peak value of these downflows reached
40~km\,s$^{-1}$. After one hour, the LOS velocities dropped from an average 
of 24~km\,s$^{-1}$ to 1--7~km\,s$^{-1}$ in only seven minutes
(Fig.~\ref{FIG11}). These findings support previous results of supersonic
downflows inferred from the \ion{He}{i} 10830~\AA\ triplet in the vicinity of
growing pores \citep{Lagg2007}.

In the present work, we mainly focused on the temporal evolution of an 
AFS as seen in the chromosphere using the \ion{He}{i} 10830~\AA\ triplet. 
In the next step we will investigate whether the plasma that drained along the legs of
the arch filaments, exhibiting supersonic downflows near the footpoints, is
detected in the photosphere.

%
%

\begin{acknowledgements}
The 1.5-meter GREGOR solar telescope was built by a German consortium under 
the leadership of the Kiepenheuer-Institut f\"ur Sonnenphysik in Freiburg with
the Leibniz-Institut f\"ur Astrophysik Potsdam, the Institut f\"ur Astrophysik
G\"ottingen, and the Max-Planck-Institut f\"ur Sonnensystemforschung in
G\"ottingen as partners, and with contributions by the Instituto de
Astrof\'{\i}sica de Canarias and the Astronomical Institute of the Academy of
Sciences of the Czech Republic. SDO HMI data are provided by the Joint Science
Operations Center -- Science Data Processing. SJGM is grateful for financial
support from the Leibniz Graduate School for Quantitative Spectroscopy in
Astrophysics, a joint project of AIP and the Institute of Physics and 
Astronomy of the University of Potsdam; SJGM and PG acknowledge the support 
of the project VEGA 2/0004/16. CD has been supported by grant DE 787/3-1 of
the German Science Foundation (DFG). MC acknowledges the support from the
Spanish Ministry of Economy and Competitiveness through the project
AYA2010-18029 (Solar Magnetism and Astrophysical Spectropolarimetry) for the
development of the instrument GRIS. This project has received funding from 
the European Research Council (ERC) under the European Union’s Horizon 2020
research and innovation programme (grant agreement No. 695075) and has been
supported by the BK21 plus program through the National Research Foundation
(NRF) funded by the Ministry of Education of Korea. This study is supported by
the European Commission's FP7 Capacities Program under the Grant Agreement
number 312495. We would like to thank Drs.\ N.\ Bello Gonz\'alez, C.\ Fischer,
and R.\ Schlichenmaier for their help during the observing campaign.
\end{acknowledgements}


\bibliographystyle{aa}
\bibliography{aa-jour,cdenker_sergio}

\end{document}